\documentclass[conference]{IEEEtran}

\usepackage{cite}
\usepackage{graphicx}
\usepackage{amsmath}
\usepackage{algorithmic}
\usepackage{url}
\usepackage{float}
\usepackage{xcolor}

\usepackage{booktabs}
\usepackage{multirow}

\usepackage{caption}
\usepackage{subcaption}

\newcommand{\aspas}[1]{{``#1''}}

\newcommand\sbar[2]{{\color{gray}\rule{\dimexpr 1cm * #1 / #2}{6pt}}}

\begin{document}

\title{Understanding the Factors that Impact the Popularity of GitHub Repositories}

\author{\IEEEauthorblockN{Hudson Borges, Andre Hora, Marco Tulio Valente}
\IEEEauthorblockA{ASERG Group, Department of Computer Science (DCC)\\
Federal University of Minas Gerais (UFMG), Brazil\\
E-mail: \{hsborges, hora, mtov\}@dcc.ufmg.br}}

\maketitle

\begin{abstract}
Software popularity is a valuable information to modern open source developers, who constantly want to know if their systems are attracting new users, if new releases are gaining acceptance, or if they are meeting user's expectations. In this paper, we describe a study on the popularity of software systems hosted at GitHub, which is the world's largest collection of open source software. GitHub provides an explicit way for users to manifest their satisfaction with a hosted repository: the stargazers button. In our study, we reveal the main factors that impact the number of stars of GitHub projects, including programming language and application domain. We also study the impact of new features on project popularity.
Finally, we identify four main patterns of popularity growth, which are derived after clustering the time series representing the number of stars of 2,279 popular GitHub repositories. We hope our results provide valuable insights to developers and maintainers, which could help them on building and evolving systems in a competitive software market.
\end{abstract}


\section{Introduction}

GitHub is the world's largest collection of open source software, with around 9 million users and 17 million public repositories.\footnote{https://github.com/search/advanced, verified on 04/04/2016} In addition to a  {\tt git}-based version control system, GitHub integrates many features for social coding. For example, developers can \emph{fork} their own copy of a repository, work and improve the code locally, and then submit a \emph{pull request} to integrate their changes in the main repository. The key characteristics and challenges of this pull-based development model is recently explored in many studies~\cite{gousios2014exploratory, Gousios2015, Yu2015, GSB16}. However, GitHub also supports other typical features from social networks. For example, users can \emph{star} a repository to manifest their interest or satisfaction with the hosted project. Consequently, the number of stars of a GitHub repository can be seen as a proxy of its popularity.
Currently, the two most popular repositories on GitHub are {\sc FreeCodeCamp/FreeCodeCamp} (a coding education software, which claims to have more than 300K users\footnote{https://www.freecodecamp.com/about, verified on 04/04/2016}) and  {\sc twbs/bootstrap} (a library of HTML and CSS templates, which is used by almost 7M web sites\footnote{http://trends.builtwith.com/docinfo/Twitter-Bootstrap, verified 04/04/2016}).



A deep understanding of the factors that impact the number of stars of GitHub repositories is important to software developers because they want to know whether their systems are attracting new users, whether the new releases are gaining acceptance, whether their systems are as popular as competitor systems, etc.
Unfortunately, we have few studies about the popularity of GitHub systems. The exceptions are
probably an attempt to differentiate popular and unpopular Python repositories using machine learning techniques~\cite{Weber2014} and a study on the effect of project's popularity on documentation quality~\cite{Aggarwal2014}. By contrast, popularity is extensively studied on other social platforms, like YouTube~\cite{Ahmed2013, Figueiredo2014} and Twitter~\cite{Lehmann2012, ma2013predicting}.
These studies are mainly conducted to guide content generators on producing successful social media content. Similarly, knowledge on software popularity might also provide valuable insights on how to build and evolve systems in a competitive market.

This paper presents an in-depth investigation on the popularity of GitHub repositories.  We first collected historical data about the number of stars of 2,500 popular repositories. We use this dataset to answer four research questions:\\[-.3cm]

\noindent \emph{RQ \#1: How popularity varies per programming language, application domain, and repository owner?} The goal is to provide an initial view about the popularity of the studied systems, by comparing the number of stars according to programming language, application domain, and repository owner (user or organization).\\[-.3cm]

\noindent \emph{RQ \#2: Does popularity correlate with other characteristics of a repository, like age, number of commits, number of contributors, and number of forks?} This investigation is important to check whether there are factors that can be worked to increase a project's popularity.\\[-.3cm]

\noindent \emph{RQ \#3: How early do repositories get popular?} With this research question, we intend to check whether gains of popularity are concentrated in specific phases of a repository's lifetime, specifically in early releases.\\[-.3cm]

\noindent \emph{RQ \#4: What is the impact of new features on popularity?} This investigation can show if relevant gains in popularity happen due to new features (implemented in new releases).\\[-.3cm]

In the second part of the paper, we identify four patterns of popularity growth in GitHub, which are derived after clustering the time series that describe the growth of the number of stars of the systems in our dataset. These patterns can help developers to understand how their systems have grown in the past and to predict future growth trends.
Finally, in the third part of the paper, we present a qualitative study with GitHub developers to clarify some findings and themes of our study.
A total of 44 developers participated to the study.

The main contribution of this paper is an investigation of factors that may impact the popularity of GitHub repositories, including the identification of the major patterns that can be used to describe popularity trends.
Although similar studies exist for social networks, to our knowledge we are the first to focus on the popularity of systems hosted in an ultra-large repository of open source code.\\[-.2cm]

\noindent \emph{Organization:} The rest of this paper is organized as follows. Section~\ref{sec:dataset} describes and characterizes the dataset used in this study. Section~\ref{sec:results}  uses this dataset to provide answers to four questions about the popularity of GitHub repositories. Section~\ref{sec-patterns} documents four patterns that describe the popularity growth of GitHub systems. Section~\ref{sec:survey}  reports the feedback of GitHub developers about three specific themes of our study. Section~\ref{sec:threats} discusses threats to validity and Section~\ref{sec:related-work} presents related work. Finally, Section~\ref{sec-conclusion} concludes the paper and lists future work.

\section{Dataset}
\label{sec:dataset}

The dataset used in this paper includes the top-2,500 public repositories with more stars in GitHub. We limit the study to 2,500 repositories for two major reasons. First,
to focus on the characteristics of the highly popular GitHub
systems. Second, because we investigate the impact of application
domain on popularity, which demands a manual classification of the
domain of each system.

All data was obtained using the GitHub API, which provides services to search public repositories and to retrieve specific data about them (e.g., stars, commits, contributors, and forks). The data was collected on March 28th, 2016.
Besides retrieving the number of stars on this date for each system, we also relied on GitHub API to collect historical data about the number of stars. For this purpose, we used a service from the API that returns all star events of a given repository. For each star, these events store the date and the user who starred the repository. However, GitHub API returns at most 100 events by request (i.e., a page) and at most 400 pages. For this reason, it is not possible to retrieve all stars events of systems with more than 40K stars, as is the case of {\sc FreeCodeCamp}, {\sc Bootstrap}, {\sc AngularJS}, {\sc D3}, and {\sc Font-Awesome}. Therefore, these five systems are not considered when answering the third and fourth research questions (that depend on historical data) and also on the study about common growth patterns (Section~\ref{sec-patterns}).

Table~\ref{tab:top10-repositories-statistics} shows descriptive statistics on the number of stars of the repositories in our dataset. The number of stars ranges from 2,150 (for {\sc CyberAgent/android-gpuimage}) to 97,948 stars (for {\sc FreeCodeCamp/FreeCodeCamp}). The median number of stars is 3,441.\\[-.25cm]

\begin{table}[!ht]
\centering
\caption{Descriptive statistics on the number of stars of the repositories in our dataset of 2,500 popular GitHub systems}
\label{tab:top10-repositories-statistics}
\begin{tabular}{@{}ccccc@{}}
\toprule
Min & 1st Quartile & 2nd Quartile & 3rd Quartile & Max \\
\midrule
2,150 & 2,682 & 3,441 & 5,331 & 97,948 \\
\bottomrule
\end{tabular}
\end{table}

\begin{figure}[!h]
\centering
\begin{subfigure}[t]{0.45\columnwidth}
\includegraphics[width=.9\columnwidth,trim={0 2em 0 0},clip]{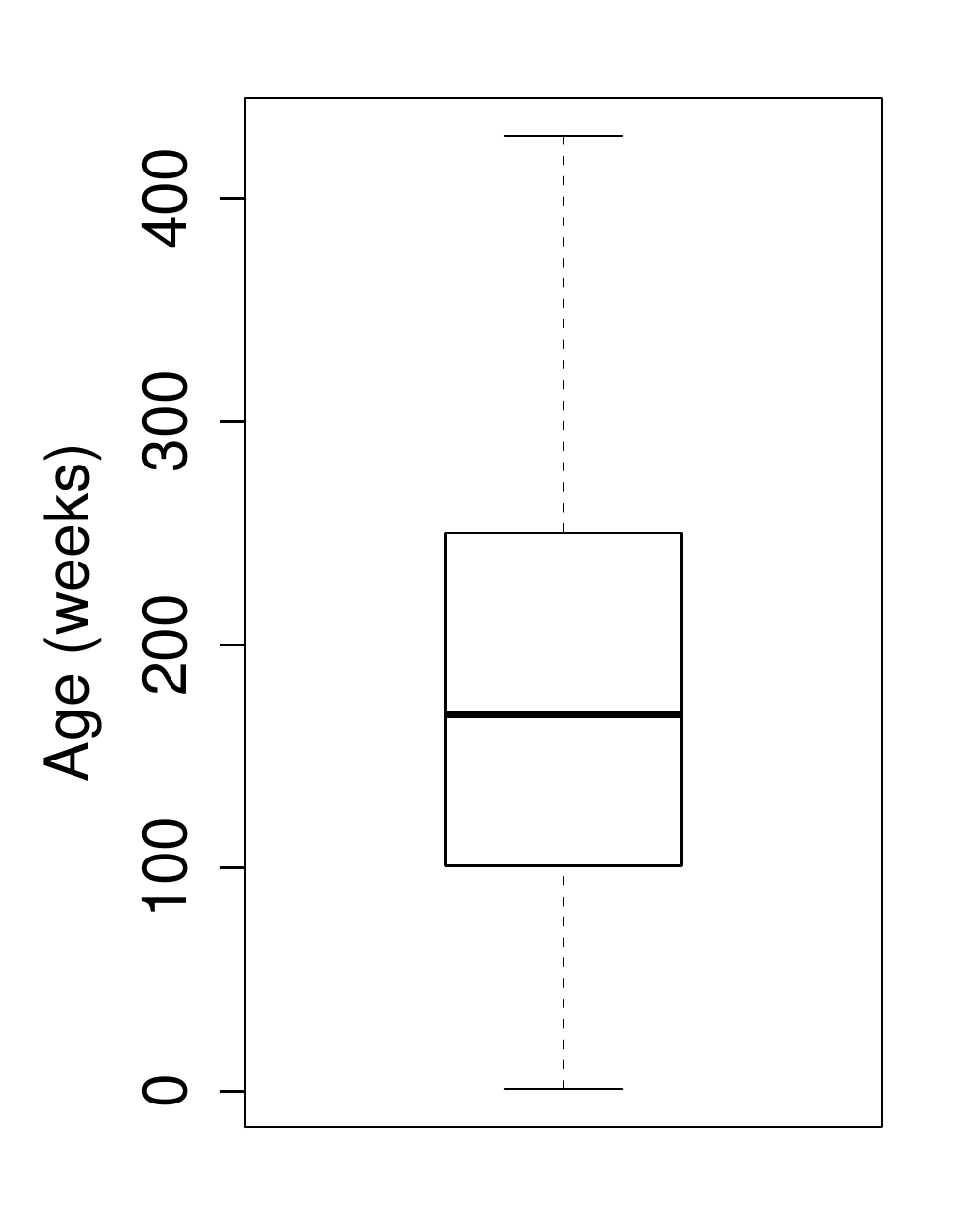}
\caption{Age (weeks)}
\label{fig:age-overview}
\end{subfigure}%
\begin{subfigure}[t]{0.45\columnwidth}
\includegraphics[width=.9\columnwidth,trim={0 2em 0 0},clip]{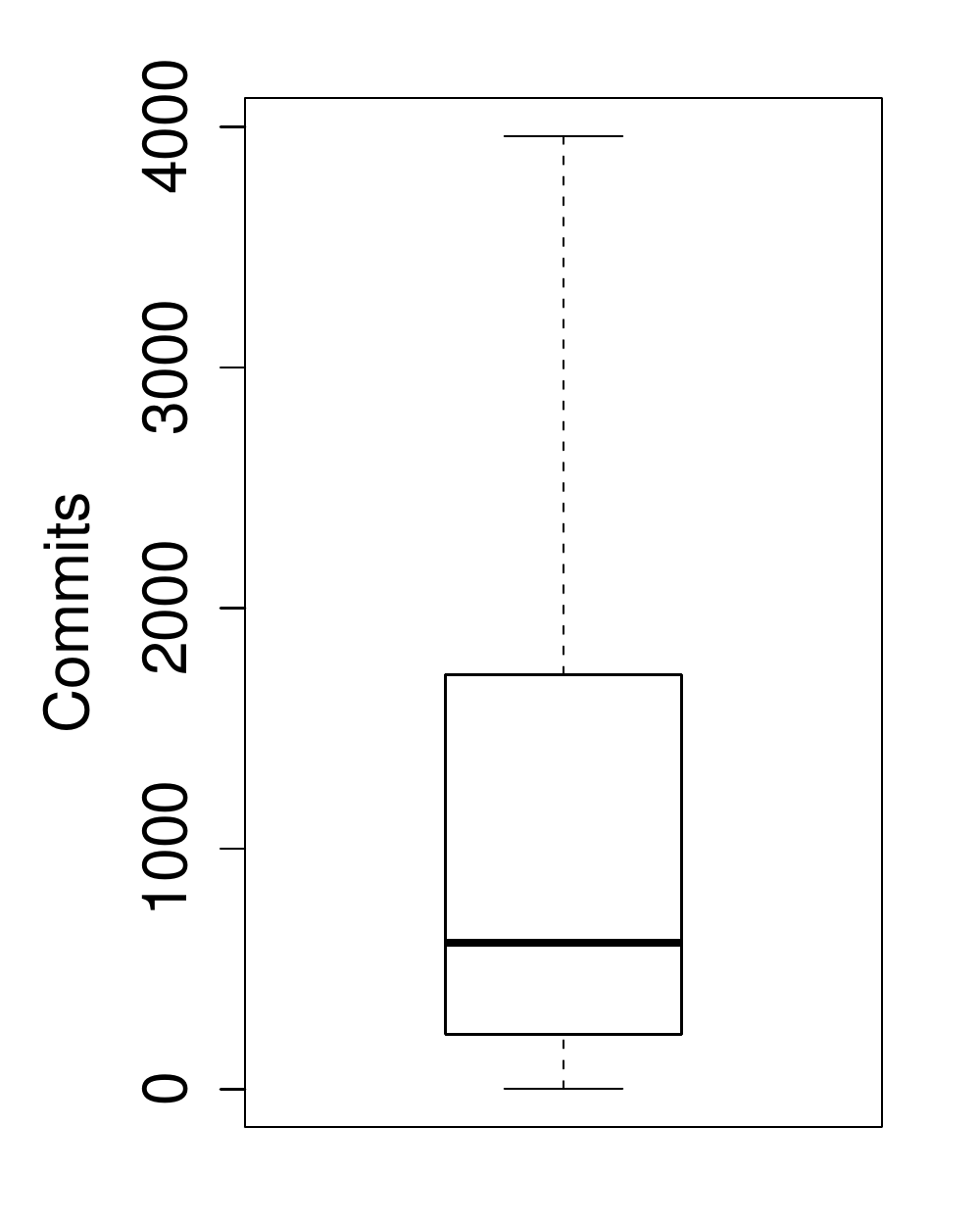}
\caption{Commits}
\label{fig:commits-overview}
\end{subfigure}

\begin{subfigure}[t]{0.45\columnwidth}
\includegraphics[width=.9\columnwidth,trim={0 2em 0 0},clip]{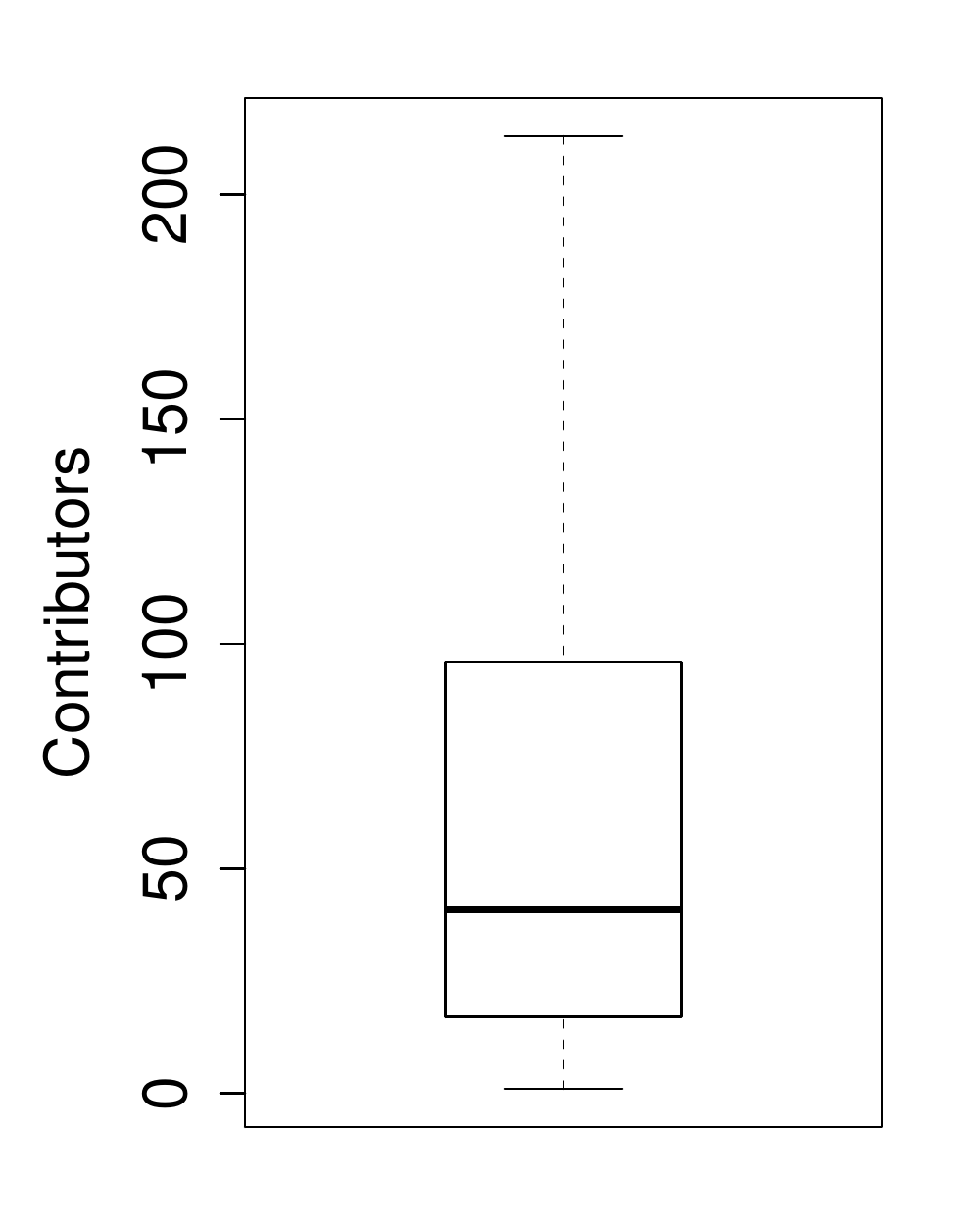}
\caption{Contributors}
\label{fig:size-overview}
\end{subfigure}%
\begin{subfigure}[t]{0.45\columnwidth}
\includegraphics[width=.9\columnwidth,trim={0 2em 0 0},clip]{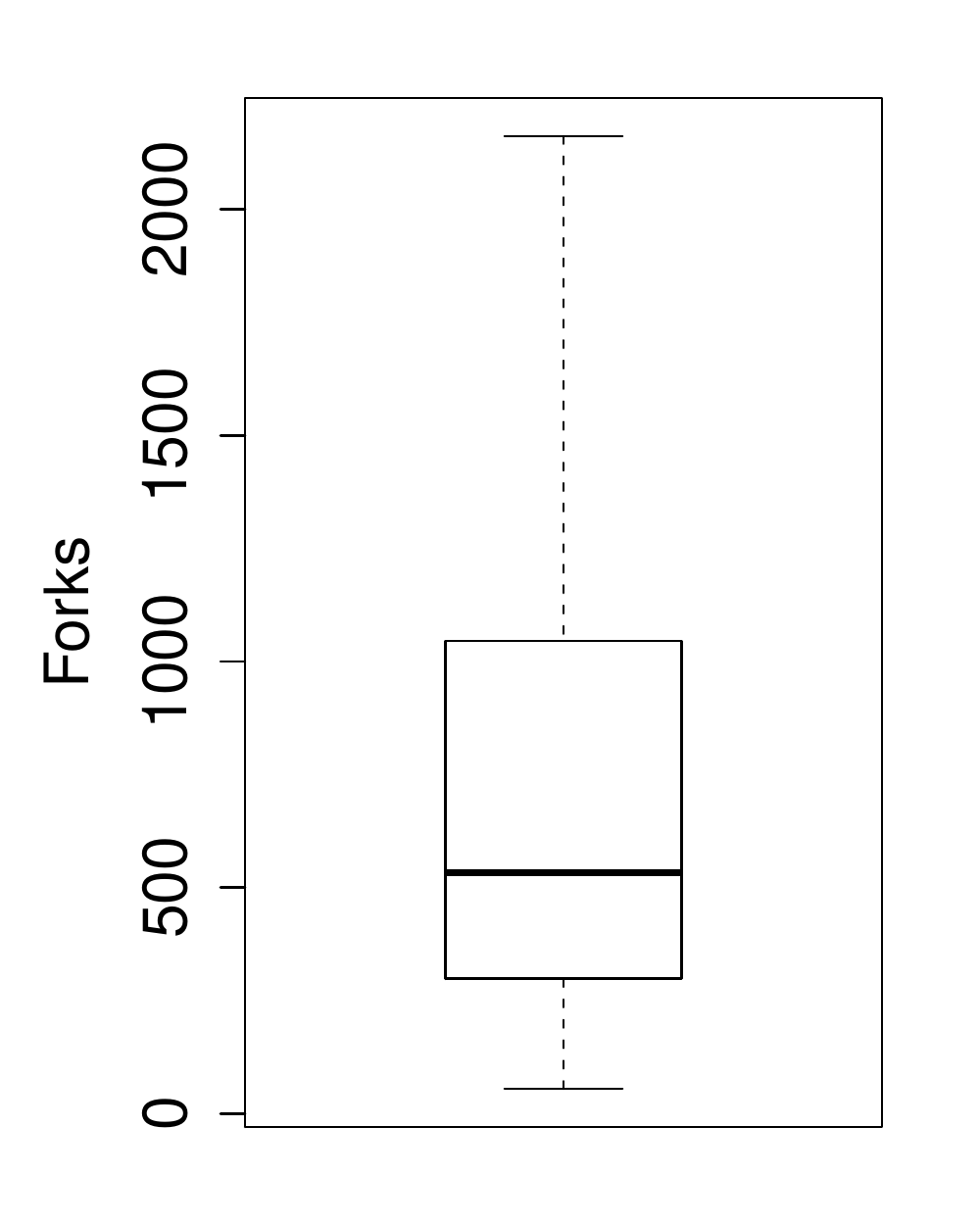}
\caption{Forks}
\label{fig:forks-overview}
\end{subfigure}
\caption{Age, number of commits, number of contributors, and number of Forks (outliers are omitted)}
\label{fig:dataset-info}
\end{figure}

\noindent \emph{Age, Commits, Contributors, and Forks:} Figure~\ref{fig:dataset-info} shows the distribution of the age (in number of weeks), number of commits, number of contributors, and number of forks for the 2,500 systems in the dataset.
For age, the first, second, and third quartiles are 101, 169, and 250 weeks, respectively.
For number of commits, the first, second, and third quartiles are 228, 608, and 1,721, respectively.
For number of contributors, the first, second, and third quartiles are 17, 41, and 96, respectively;\footnote{We report data from contributors as retrieved by GitHub API. This data may be different from the one presented on the project's page at GitHub, which only counts contributors with GitHub account.} and for number of forks, the first, second, and third quartiles are 298, 533, and 1,045, respectively. Therefore, the systems in our dataset are mature and have many commits and contributors.\\[-.25cm]

\begin{figure}[!h]
\centering
\includegraphics[width=1\columnwidth, trim={0 1em 0 4em}, clip]{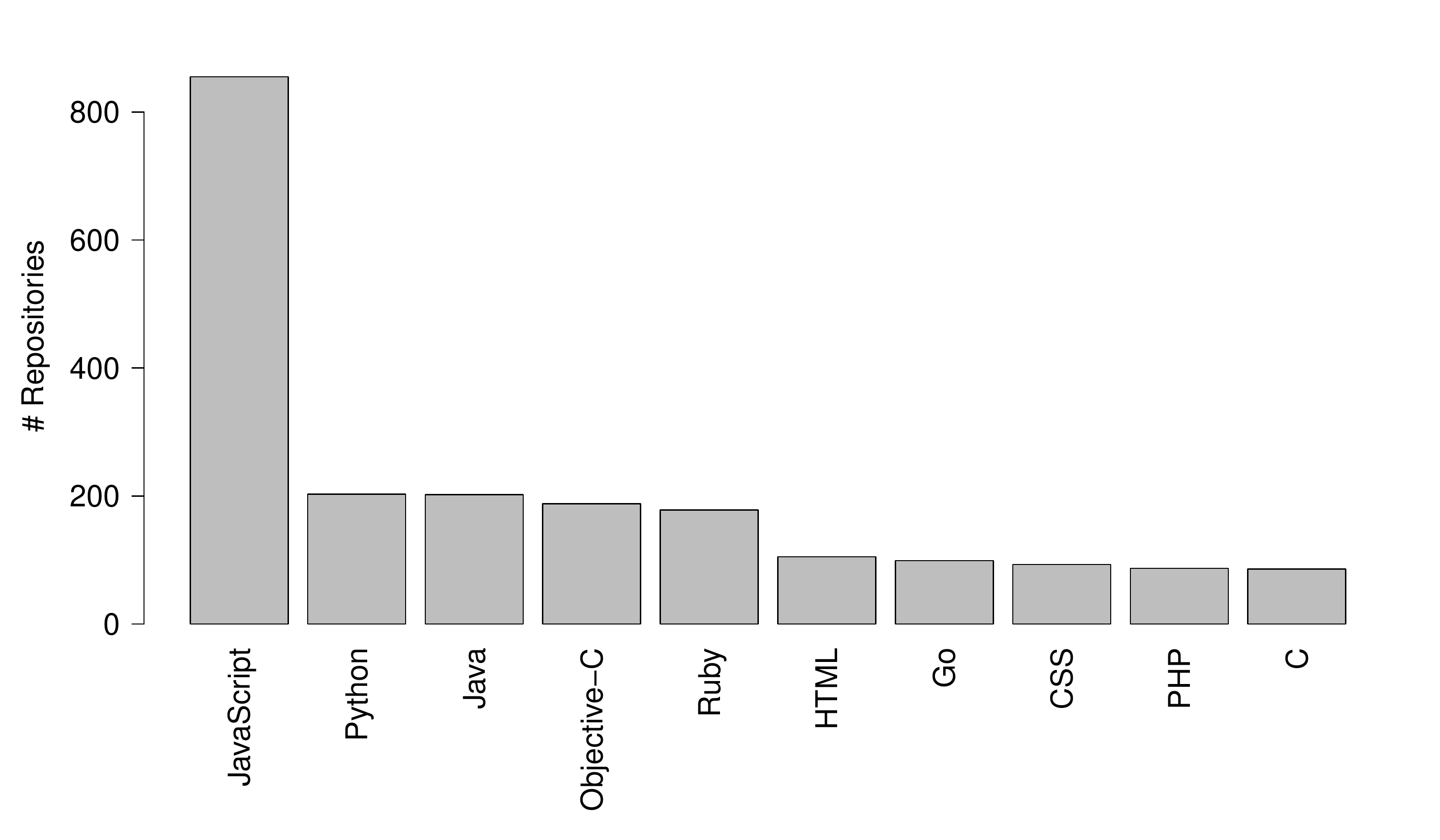}
\caption{Top-10 languages by number of repositories}
\label{fig:stars-overview2}
\end{figure}

\noindent\emph{Programming Language:} As returned by GitHub API, the language of a repository is the one with the highest percentage of source code, considering the files in the repository. Figure~\ref{fig:stars-overview2} shows the distribution of the systems per programming language. JavaScript is the most popular language (855 repositories, 34.2\%), followed by Python (203 repositories, 8.1\%), Java (202 repositories, 8.0\%), Objective-C (188 repositories, 7.5\%), and Ruby (178 repositories, 7.1\%). Despite a concentration of systems in these languages, the dataset includes systems in 53 languages, including Groovy, R, Julia, and XSLT (all with just one repository).\\[-.25cm]

\noindent\emph{Owner:} We also provide results grouped by repository owner. In GitHub, a repository can be owned by a user (e.g.,~{\sc torvalds/linux}) or by an organization (e.g.,~{\sc facebook/react}). In our dataset, 1,263 repositories (50.5\%) are owed by organizations and 1,237 repositories (49.5\%) by users.\\[-.25cm]

\noindent\emph{Application Domain:} In the study reported in this paper, results are also grouped by application domain. However, different from other source code repositories, like SourceForge, GitHub does not include information about the application domain of a project. For this reason, we manually classified the domain of each system in our dataset. Initially, the first and third authors of this paper inspected the description of the top-200 repositories to provide a first list of application domains. After this initial classification, the first author inspected the short description (and in many cases the GitHub page and the project's page) of the remaining 2,300 repositories. During this process, he also marked the repositories with dubious classification decisions. These particular cases were discussed by the first and second authors, to reach a consensus decision. The spreadsheet with the proposed classification is publicly available at https://goo.gl/73Sbvz.

The systems are classified in the following six domains:

\begin{itemize}

\item Application software: systems that provide functionalities to end-users, like browsers and text editors (e.g.,~{\sc WordPress/WordPress} and {\sc adobe/brackets}).
\item System software: systems that provide services and infrastructure to other systems, like operating systems, middleware, servers, and databases (e.g.,~{\sc torvalds/linux} and {\sc mongodb/mongo}).
\item Web libraries and frameworks (e.g.,~{\sc twbs/bootstrap} and {\sc angular/angular.js}).
\item Non-web libraries and frameworks (e.g.,~{\sc google/guava} and {\sc facebook/fresco}).
\item Software tools: systems that support software development tasks, like IDEs, package managers, and compilers (e.g.,~{\sc Homebrew/homebrew} and {\sc git/git}).
\item Documentation: repositories with documentation, tutorials, source code examples, etc. (e.g.,~{\sc iluwatar/java-design-patterns}).
\end{itemize}

\begin{figure}[!h]
\centering
\includegraphics[width=.8\columnwidth, trim={0 0 0 3em}, clip]{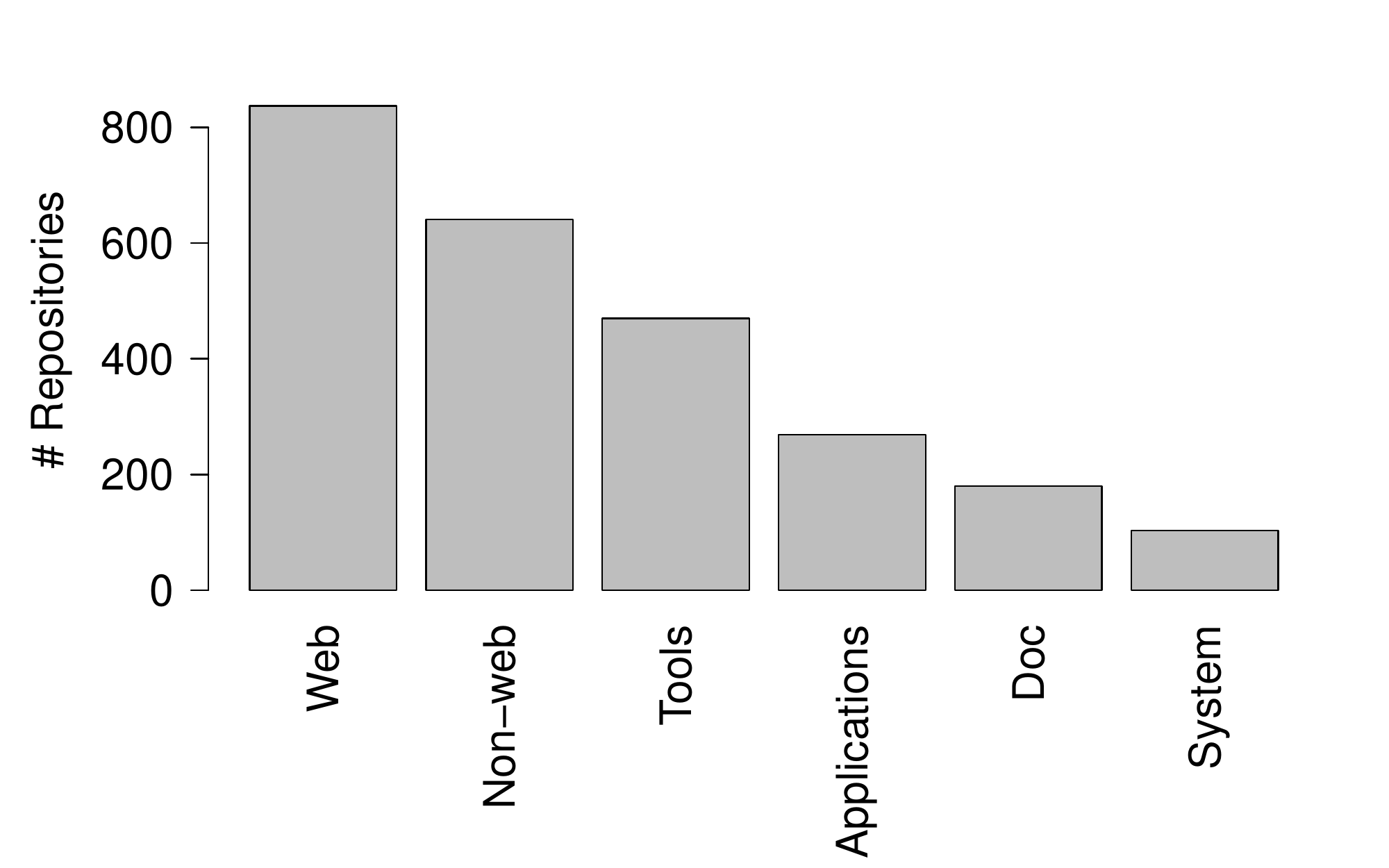}
\caption{Number of repositories by domain}
\label{fig:domains-dist}
\end{figure}

Figure~\ref{fig:domains-dist} shows the number of systems in each domain. The top-3 domains are web libraries and frameworks (837 repositories, 33\%), non-web libraries and frameworks (641 repositories, 25\%), and software tools (470 repositories, 18\%).

\section{Results}
\label{sec:results}

In this section, we use the described dataset to answer the four research questions listed in the paper's introduction.

\vspace{1em}\noindent\emph{\textbf{RQ \#1}: How popularity varies per programming language, application domain, and repository owner?}\vspace{0.5em}
\label{sub:results:rq2}

\begin{figure}[!h]
\centering
\includegraphics[width=.95\columnwidth, trim={0 0 0 2em}, clip]{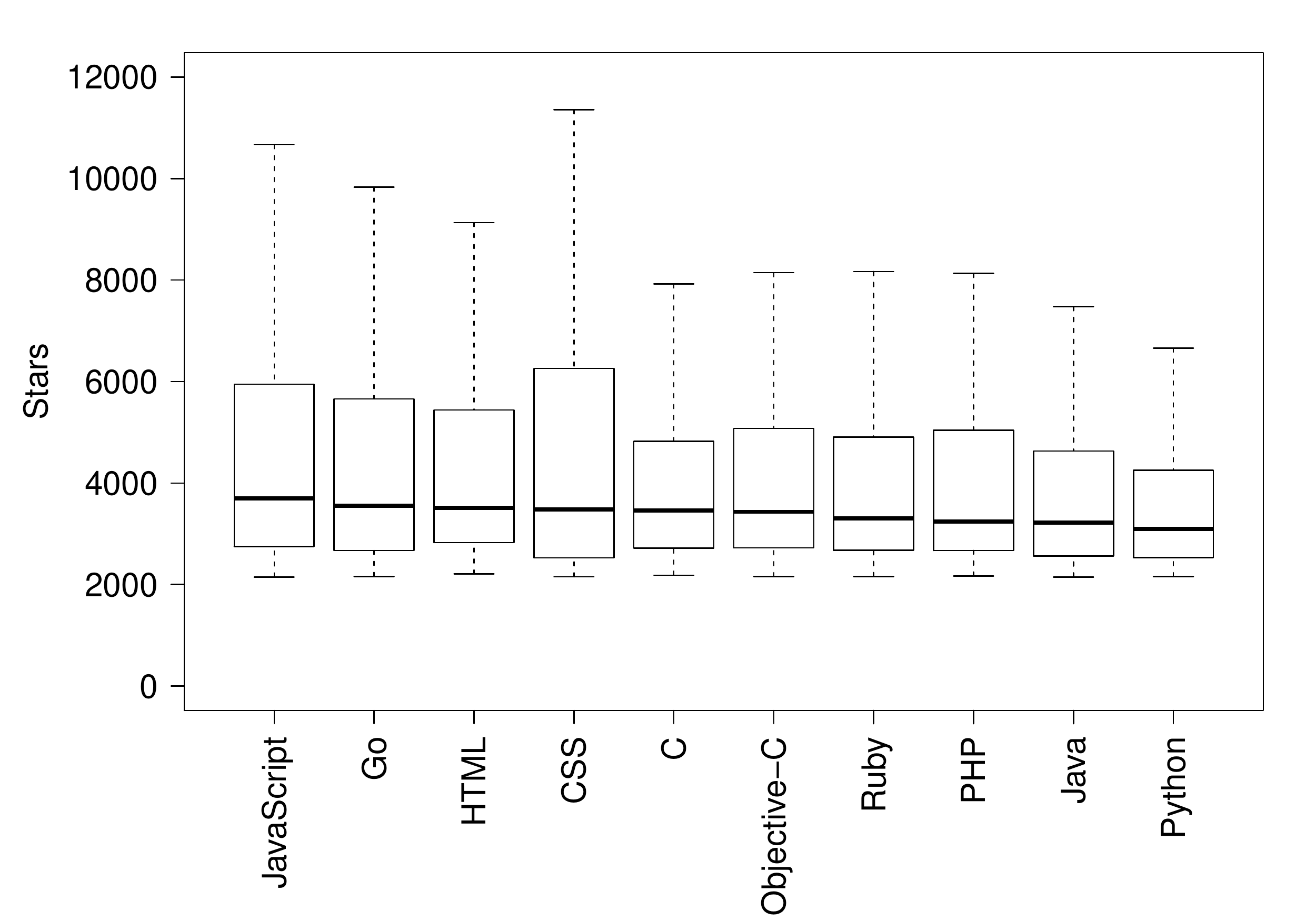}
\caption{Stars by programming language (considering only the top-10 languages with more repositories)}
\label{fig:language-popularity}
\end{figure}

Figure~\ref{fig:language-popularity} shows the distribution of the number of stars for the top-10 languages with more repositories.
The top-3 languages whose repositories have the highest median number of stars are: JavaScript (3,697 stars), Go (3,549 stars), and HTML (3,513 stars).
The three languages whose repositories have the lowest median number of stars are PHP (3,245 stars), Java (3,224 stars), and Python (3,099 stars).
By applying the Kruskal-Wallis test to compare multiple samples, we find that the distribution of the number of stars per language is different (\emph{p-value} $=$ 0.001).
Thus, we can consider that programming language may impact on system popularity.

\begin{figure}[!h]
\centering
\includegraphics[width=.95\columnwidth, trim={0 0 0 1em}, clip]{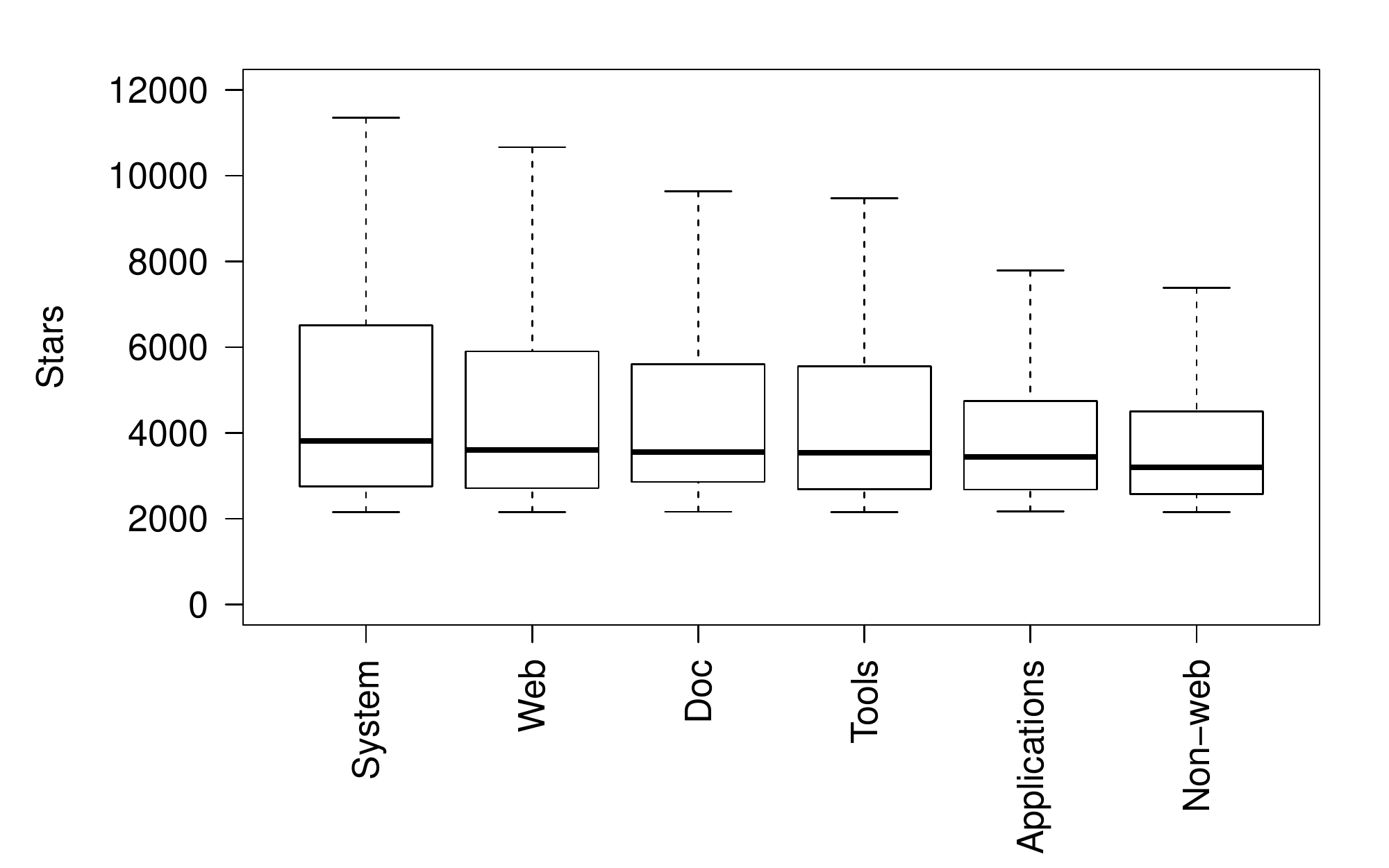}
\caption{Popularity by application domain}
\label{fig:domains-popularity}
\end{figure}

Figure~\ref{fig:domains-popularity} shows the distribution of the number of stars for the repositories in each application domain. The median number of stars varies as follow: systems software (3,807 stars), web libraries and frameworks (3,596 stars), documentation (3,547 stars), software tools (3,538 stars), applications (3,443 stars), and now-web libraries and frameworks (3,204 stars).
By applying the Kruskal-Wallis test, we find that the distribution of the number of stars by domain is different (\emph{p-value} $<$ 0.001).
Therefore, application domain is also an important factor that may impact on system popularity.
\\[-0.3cm]

\begin{figure}[!ht]
\centering
\includegraphics[width=.65\columnwidth, trim={0 0 0 2.5em}, clip]{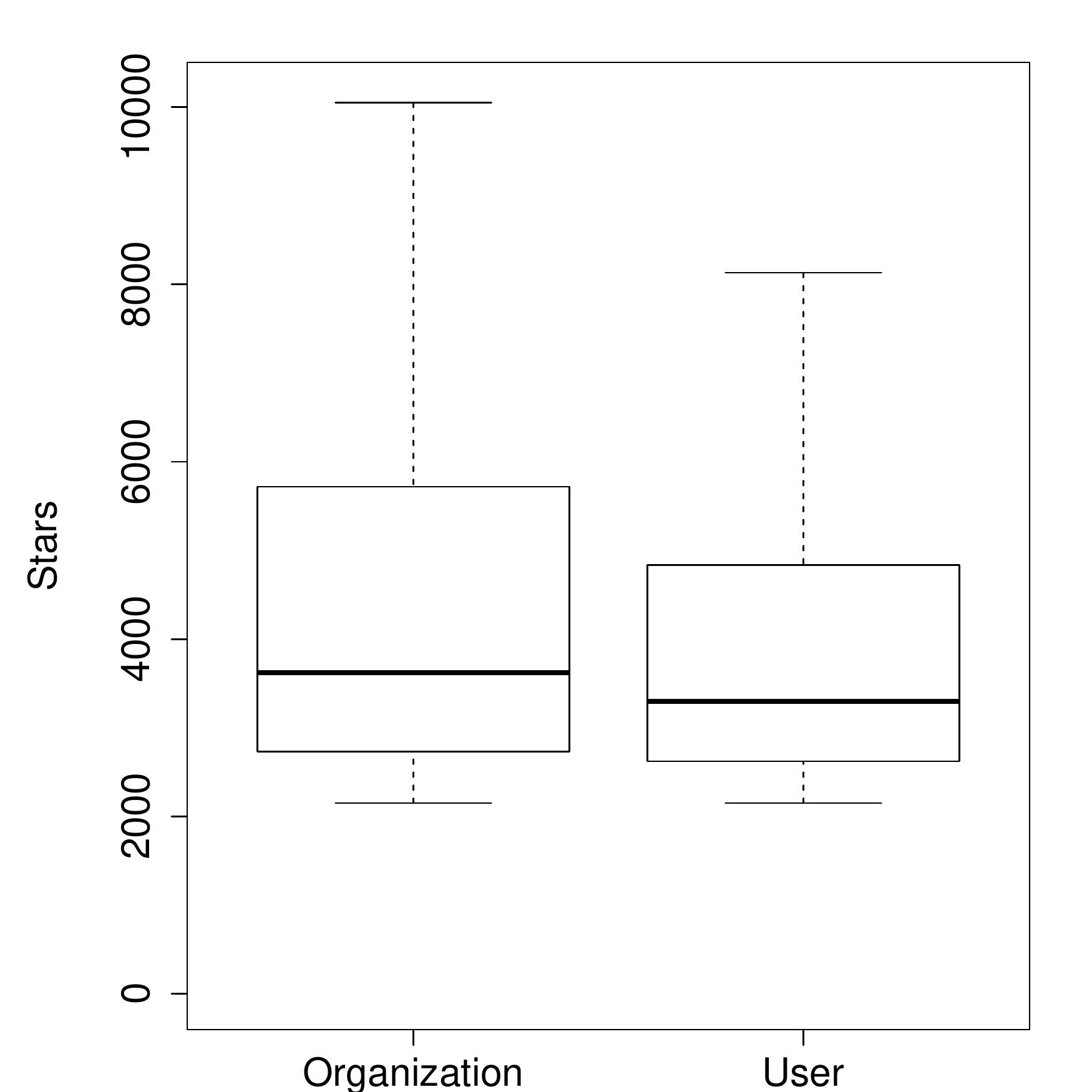}
\caption{Popularity by repository owner}
\label{fig:owner-popularity}
\end{figure}

Finally, Figure~\ref{fig:owner-popularity} shows how popularity varies depending on the repository owner (i.e., user or organization). The median number of stars is 3,622 stars for repositories owned by organizations and 3,298 stars for repositories owned by users.
By applying the Mann-Whitney test, we detect that indeed these distributions are different (\emph{p-value} $<$ 0.001).
We hypothesize that repositories owned by organizations---specially major software companies and free software
foundations---have more funding and resources, which somehow explains their higher popularity.
\\[-0.3cm]

\noindent{\em Summary: The top-5 languages with more stars are JavaScript, Python, Java, Objective-C, and Ruby (Figure~\ref{fig:stars-overview2}). However, the top-5 languages whose systems have the highest median number of stars are JavaScript, Go, HTML, CSS, and C (Figure~\ref{fig:language-popularity}). The top-3 application domains whose repositories have more stars are systems software, web libraries and frameworks, and documentation. Repositories owned by organizations are more popular than the ones owned by individuals.}

\begin{figure*}[!t]
\centering
\begin{subfigure}[b]{0.25\textwidth}
\includegraphics[width=\linewidth, trim={0 1em 0 4.5em}, clip]{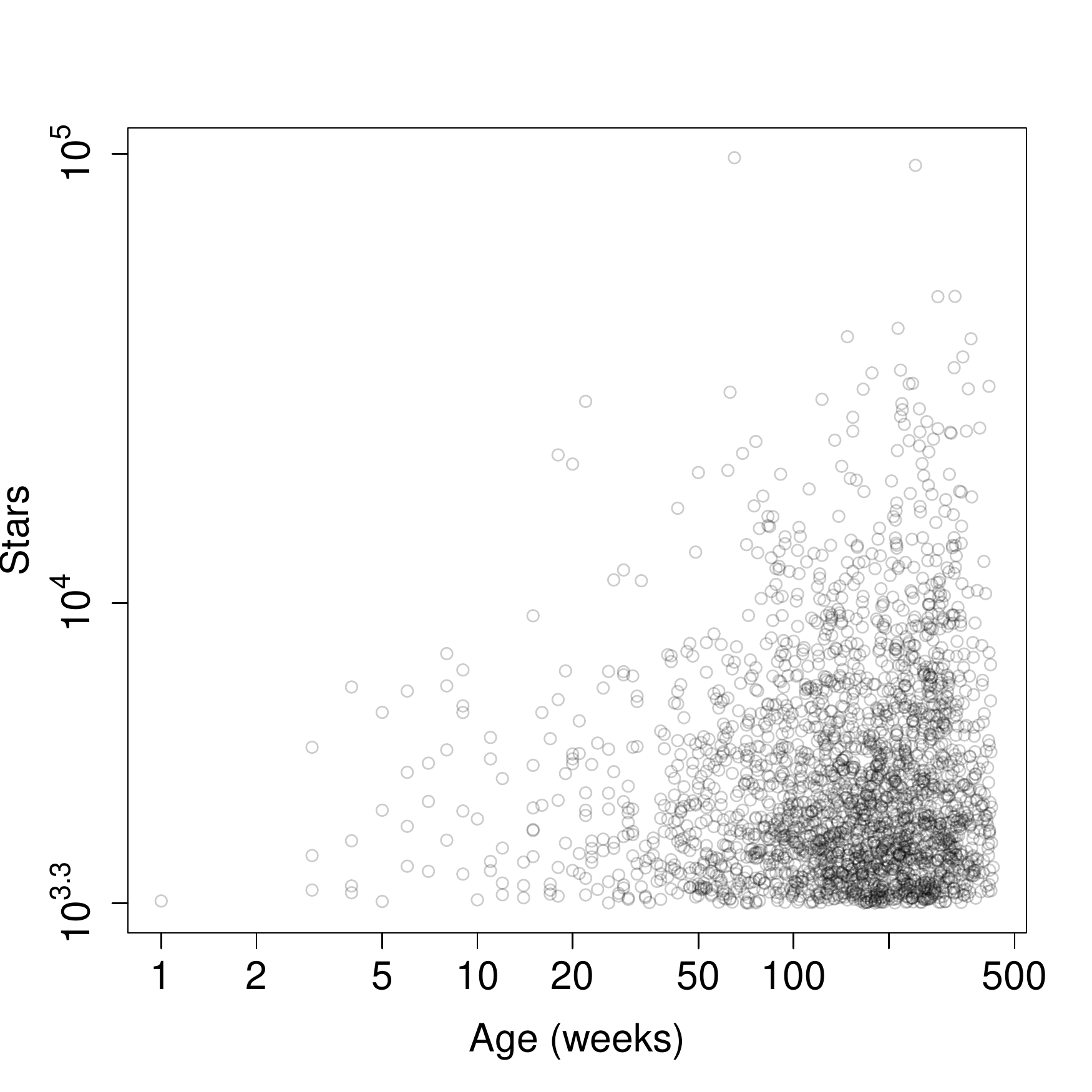}
\caption{Age vs Stars}
\label{fig:age-vs-stars}
\end{subfigure}%
\begin{subfigure}[b]{0.25\textwidth}
\includegraphics[width=\linewidth, trim={0 1em 0 4.5em}, clip]{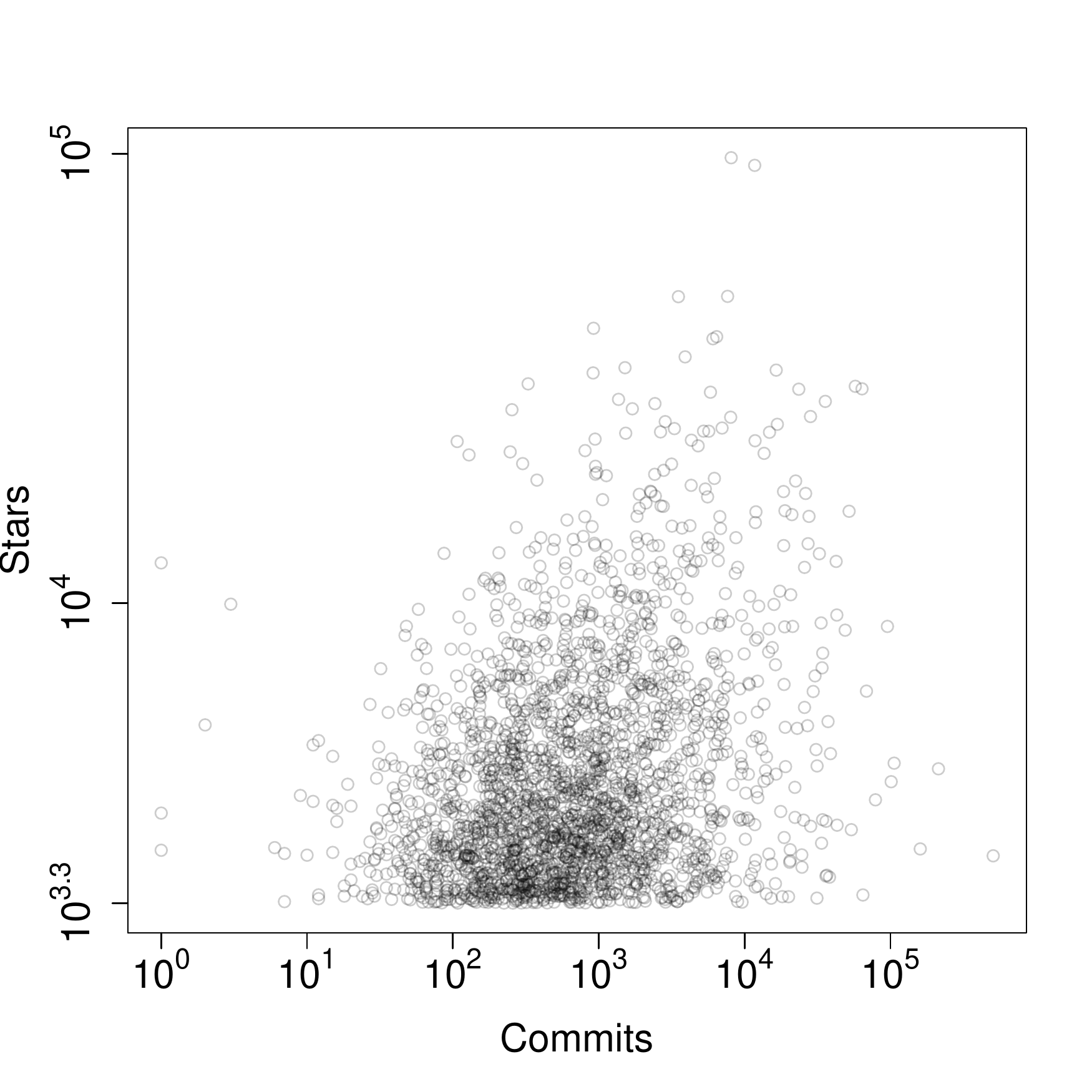}
\caption{Commits vs Stars}
\label{fig:commits-vs-stars}
\end{subfigure}%
\begin{subfigure}[b]{0.25\textwidth}
\includegraphics[width=\linewidth, trim={0 1em 0 4.5em}, clip]{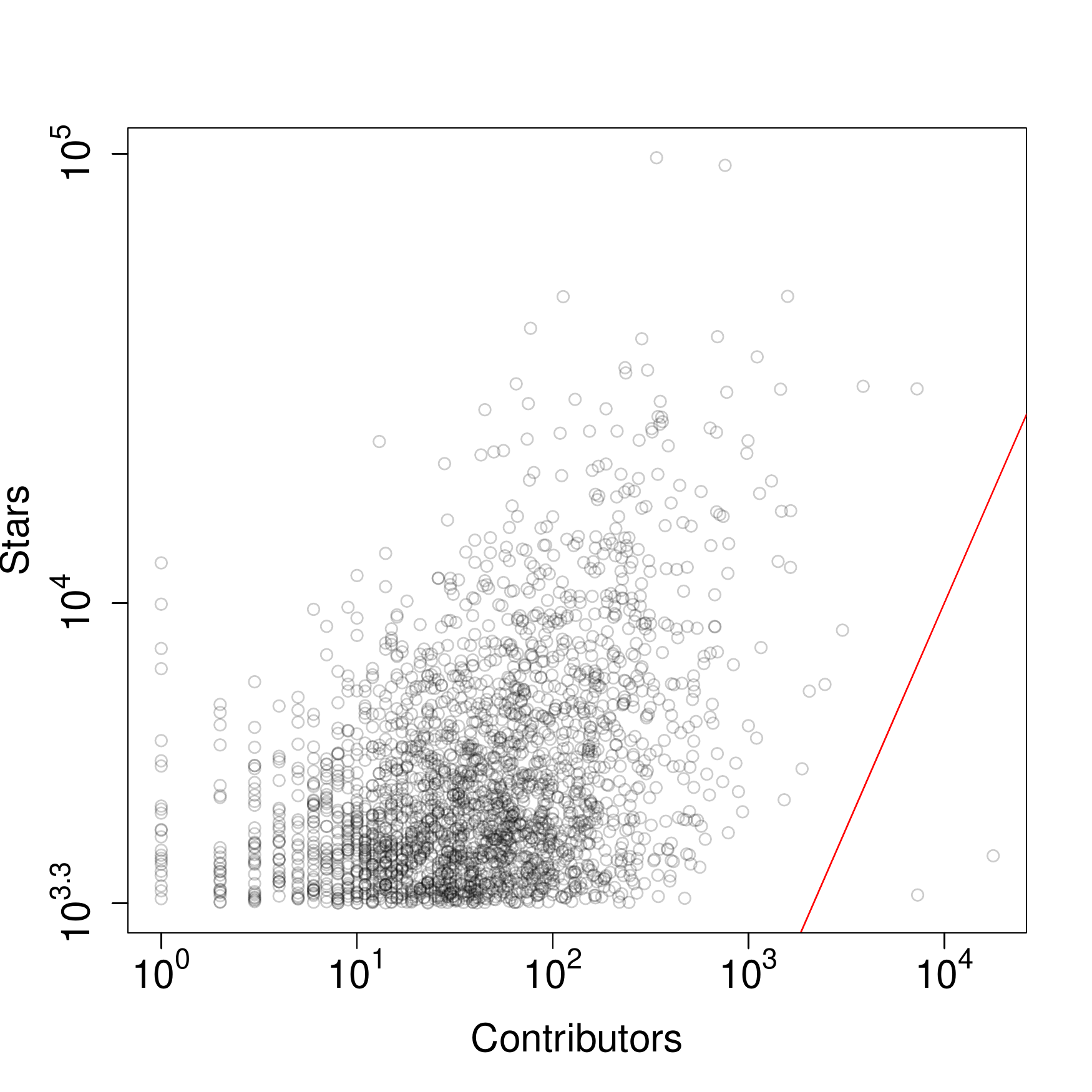}
\caption{Contributors vs Stars}
\label{fig:contributors-vs-stars}
\end{subfigure}%
\begin{subfigure}[b]{0.25\textwidth}
\includegraphics[width=\linewidth, trim={0 1em 0 4.5em}, clip]{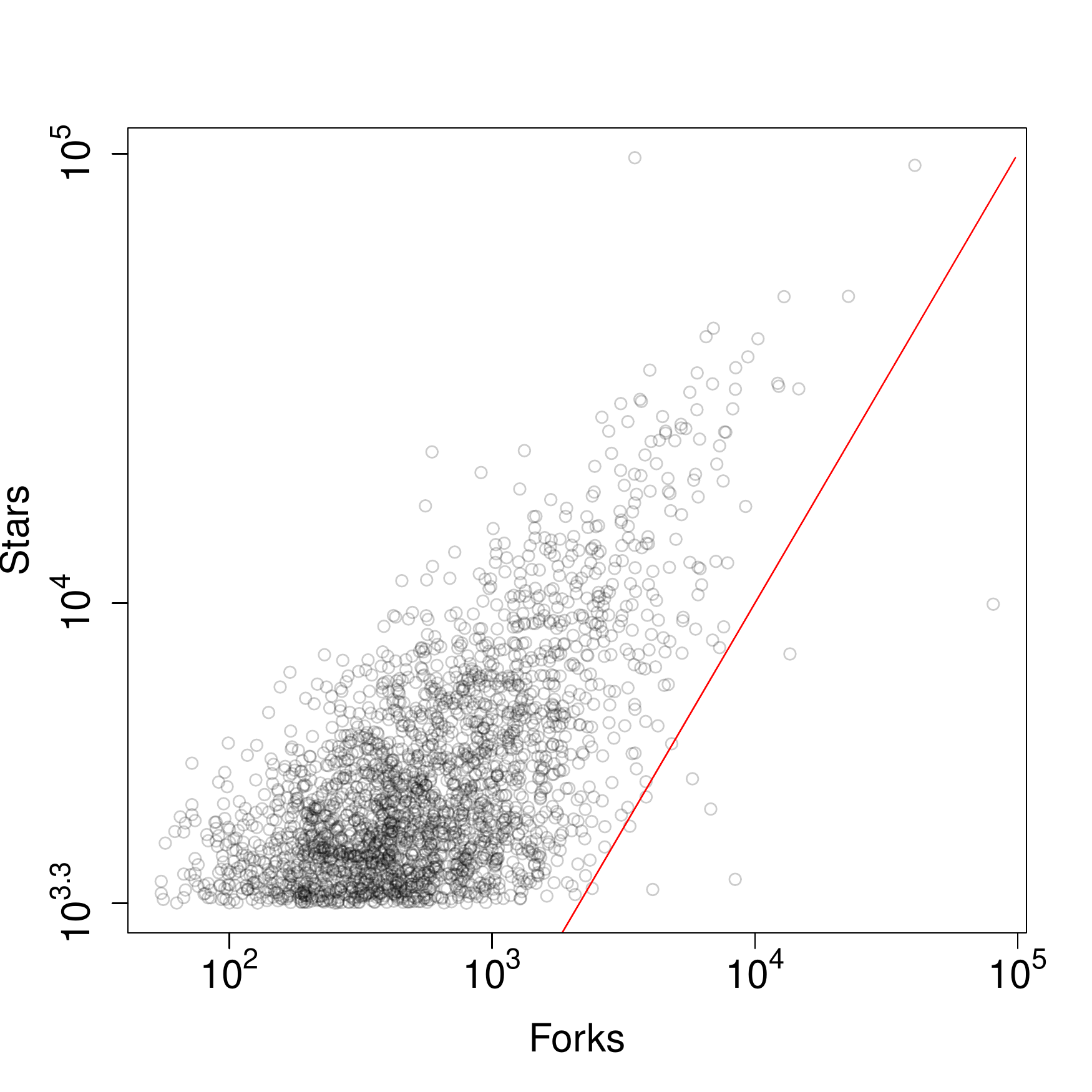}
\caption{Forks vs Stars}
\label{fig:forks-vs-stars}
\end{subfigure}
\caption{Correlation analysis. In subfigures (c) and (d), the line is the identity relation}
\label{fig:correlations}
\end{figure*}

\vspace{1em}\noindent\emph{\textbf{RQ \#2}: Does popularity correlate with repository's age, number of commits, number of contributors, and number of forks?}\vspace{0.5em}
\label{sub:results:rq3}

Figure~\ref{fig:correlations} shows scatterplots correlating the number of stars with the age (in number of weeks), number of commits, number of contributors, and number of forks of a repository.
First, the plots suggest that stars are not correlated with the repository's age (Figure~\ref{fig:age-vs-stars}). We have old repositories with few stars and new repositories with many stars. For example, {\sc apple/swift} has only five months and 28,105 stars, while {\sc mojombo/chronic} has more than 8 years and 2,440 stars. Essentially, this result shows that repositories gain stars at different speeds. We ran Spearman's rank correlation test and the resulting correlation coefficient is close to zero ($rho$ $=$ 0.0757 and \emph{p-value} $<$ 0.001).

The scatterplot in Figure~\ref{fig:commits-vs-stars} suggests that stars are weakly correlated with number of commits ($rho$ $=$ 0.249 with \emph{p-value} $<$ 0.001). Similarly, as presented in Figure~\ref{fig:contributors-vs-stars} stars are weakly correlated with contributors ($rho$ $=$ 0.341 with \emph{p-value} $<$ 0.001).
In this figure, a logarithm scale is used in both axes; the line represents the identity relation: below the line are the systems with more contributors than stars.
Interestingly, two systems indeed have more contributors than stars: {\sc raspberrypi/linux} (17,766 contributors and 2,739 stars) and {\sc Linuxbrew/linuxbrew} (7,304 contributors and 2,241 stars).
This happens because they are forks of highly successful repositories ({\sc torvalds/linux} and {\sc Homebrew/brew}, respectively).
The top-3 systems with more stars per contributor are {\sc shadowsocks/shadowsocks} (12,287 stars/contributor), {\sc octocat/Spoon-Knife} (9,944 stars/contributor), and {\sc wg/wrk} (7,923 stars/contributor). All these systems have just one contributor. 
The three systems with less stars per contributor are {\sc android/platform\_frameworks\_base} (2.28 stars/contributor), {\sc FFmpeg/FFmpeg} (2.39 stars/contributor), and {\sc DefinitelyTyped/DefinitelyTyped} (2.68 stars/ contributor).

Finally, Figure~\ref{fig:forks-vs-stars} shows plots correlating a system popularity and its number of forks.
As visually suggested by the figure, there is a
strong positive correlation between stars and forks ($rho$ $=$ 0.549 and \emph{p-value} $<$ 0.001). For example, {\sc twbs/bootstrap}
is the second repository with the highest number of stars and the second one with more forks. {\sc angular/angular.js} is the third repository
in number of stars and the third one with more forks. In Figure~\ref{fig:forks-vs-stars}, we can also see that
only nine systems (0.36\%) have more forks than stars. As examples, we have a repository that just
provides a tutorial for forking a repository ({\sc octocat/SpoonKnife}) and a
popular puzzle game ({\sc gabrielecirulli/2048}), whose success motivated many forks with
variations of the original implementation. Since the game can be downloaded directly from
the web, we hypothesize that it receives most users' feedback in the web and not on GitHub.
 \\[-0.1cm]

\noindent {\em Summary: There is no correlation between numbers of stars and the repository's age; however, there is a weak correlation with commits and contributors. Moreover, a strong correlation with forks was found.}\\[-0.2cm]

\vspace{1em}\noindent\emph{\textbf{RQ \#3}: How early do repositories get popular?}\vspace{1em}
\label{sub:results:rq4}

\begin{figure}[!ht]
\centering
\includegraphics[width=0.85\columnwidth, page=1, trim={0 0 0 2.5em}, clip]{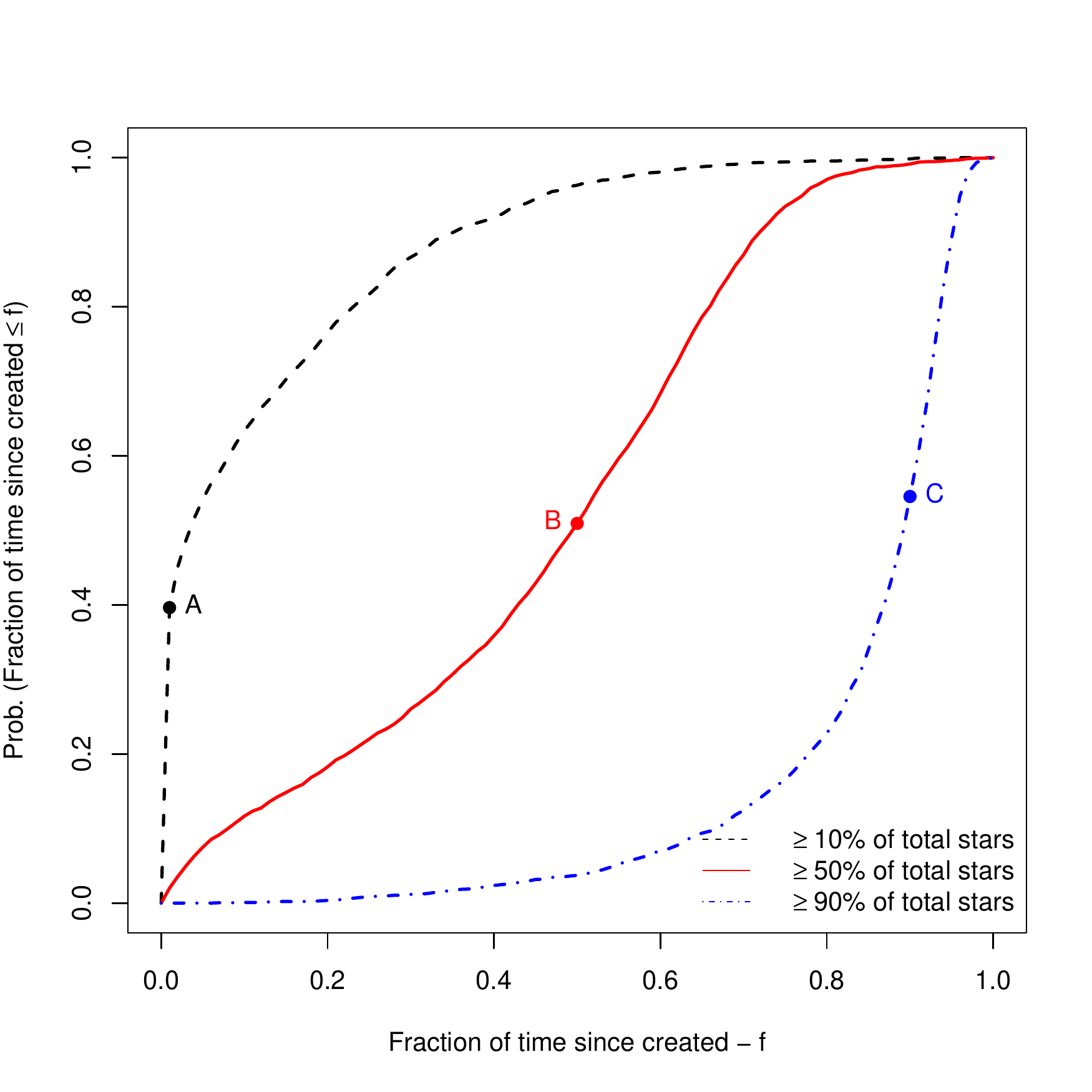}
\caption{Cumulative distribution of the fraction of time a repository takes to receive 10\%, 50\%, and 90\% of its stars}
\label{fig:cdf}
\end{figure}

Figure~\ref{fig:cdf} shows the cumulative distribution of the fraction of time a repository takes to receive at least 10\%, at least 50\%, and at least 90\% of its stars.
Specifically, the y-axis shows the fraction of repositories that achieved 10\%, 50\%, and 90\% of their stars in a period of time that does not exceed the fraction of time shown in the x-axis. Around 40\% of the repositories receive 10\% of their stars very early, in the first days after the initial release (label A, in Figure~\ref{fig:cdf}). We hypothesize that many of these initial stars come from early adopters, who start commenting and using novel open source software immediately after they are out. After this initial burst of popularity, the growth of half of the repositories tend to stabilize. For example, half of the repositories take 51\% of their age to receive 50\% of their stars (label B); and half of the repositories take 91\% of their age to receive 90\% of their total number of stars (label C). \\[-0.1cm]

{\noindent {\em Summary: Repositories have a tendency to receive more stars right after their first public release. After this period, for half of the repositories the growth rate tends to stabilize.}

\vspace{1em}\noindent\emph{\textbf{RQ \#4}: What is the impact of new features on popularity?}\vspace{1em}
\label{sub:results:rq5}

In this research question, we investigate the impact of new features on the popularity of GitHub repositories.
The goal is to check whether the implementation of new features (resulting in new releases of the projects) contribute to a boost in popularity. Specifically, we selected 834 repositories from our dataset (33.3\%) that follow a semantic versioning convention to number releases.\footnote{http://semver.org}
In such systems, versions are identified by three integers, in the format $x.y.z$, with the following semantics: increments in ${x}$ denote major releases, which can be incompatible with old versions; increments in $y$ denote minor releases, which add functionality in a backwards-compatible manner; and increments in $z$ denote patches implementing bug fixes.
In our sample, we identified 580 major releases and 4,343 minor releases.

First, we counted the fraction of stars received by each repository in the week following all releases (major or minor) and just after major releases. As mentioned, the goal is to check the impact of new releases in the number of stars.
Figure~\ref{fig:releases1} shows the distribution of these fractions.
When considering all releases, the fraction of stars gained in the first week after the releases is 1.1\% (first quartile), 3.2\% (second quartile), and 10.2\% (third quartile). For the major releases only, it is 0.5\% (first quartile), 1.4\% (second quartile), and 4.3\% (third quartile).
{\sc so-fancy/diff-so-fancy} (a visualization for git diffs) is the repository with the highest fraction of stars received after releases. The repository has 53 days and 4,402 stars. Since it has a fast releasing rate (one new release per week, on average), it gained almost of its stars (89.1\%) in the weeks after releases.

\begin{figure}
\centering
\includegraphics[width=0.65\columnwidth, page=1, trim={0 0 0 2.5em}, clip]{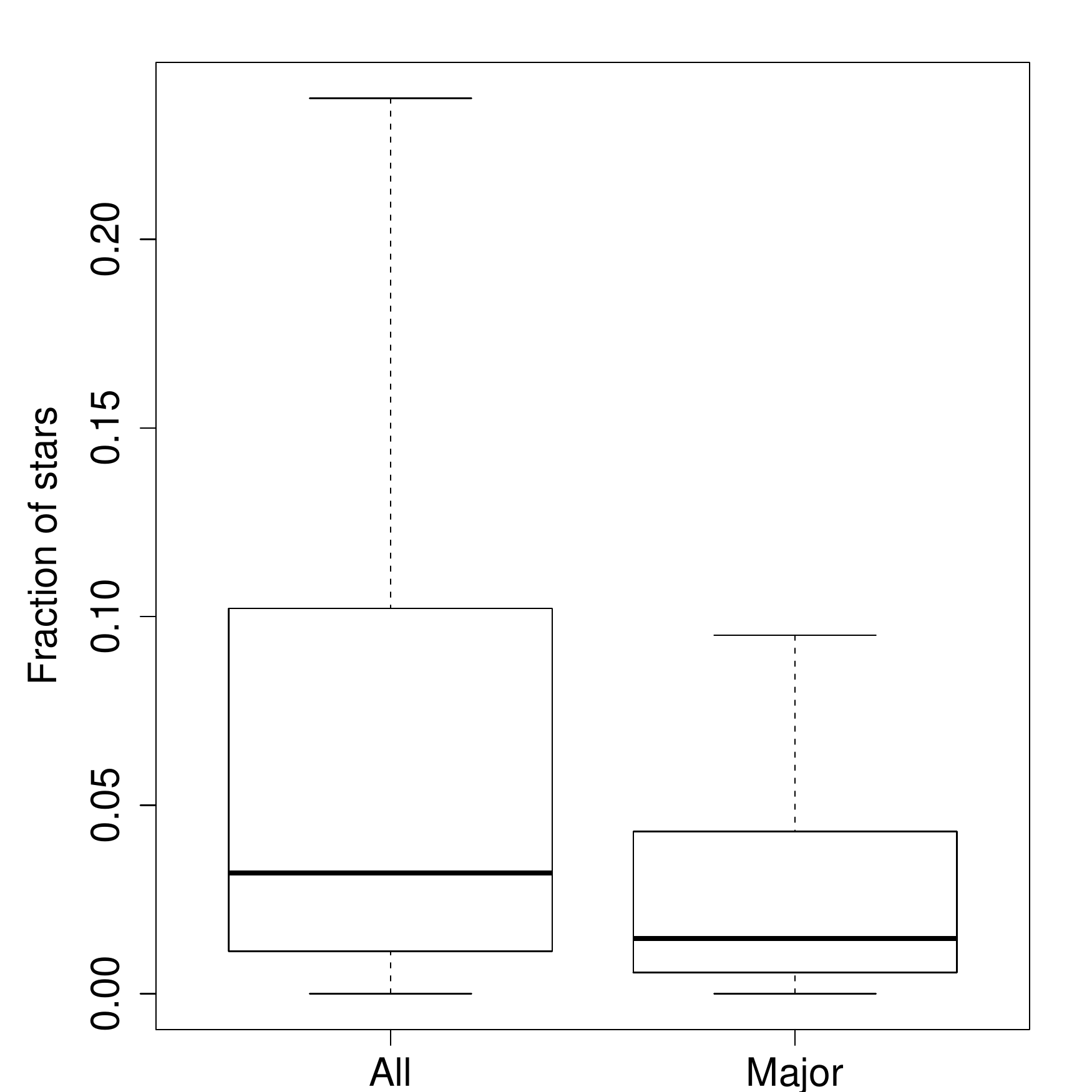}
\caption{Fraction of stars gained in the first week after all releases and just after the major releases}
\label{fig:releases1}
\end{figure}

We computed a second ratio for each repository: \emph{fraction of stars in the week following all releases or just major releases (FS)} $/$ {\em fraction of time represented by these weeks (FT)}.
When $\mathit{FS}/\mathit{FT} > 1$, the repository gains proportionally more stars after the releases.
Figure~\ref{fig:releases2} shows boxplots with the results of $\mathit{FS}/\mathit{FT}$
for all repositories.
When considering all releases, we have that $\mathit{FS}/\mathit{FT}$ is 0.80 (first quartile), 1.25 (second quartile), and 1.98 (third quartile). For major releases only, we have that $\mathit{FS}/\mathit{FT}$ is 0.81 (first quartile), 1.53 (second quartile), and 2.98 (third quartile).

\begin{figure}[!ht]
\centering
\includegraphics[width=0.65\columnwidth, page=3, trim={0 0 0 2.5em}, clip]{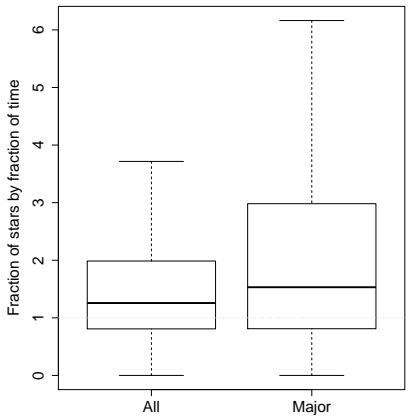}
\caption{Fraction of stars in the week following all releases (or just the major releases) $/$ fraction of time represented by these weeks}
\label{fig:releases2}
\end{figure}

Figure~\ref{fig:releases-values} shows the median values of $\mathit{FS}/\mathit{FT}$  computed using stars gained after $n$ weeks ($1 \leq n \leq 4$). This ratio decreases, both for major and for all releases. Therefore, although there is some gains of stars after releases, they tend to decrease after few weeks.\\[-0.1cm]

\begin{figure}[!h]
\centering
\includegraphics[width=0.65\columnwidth, trim={0 0 0 2.5em}, clip]{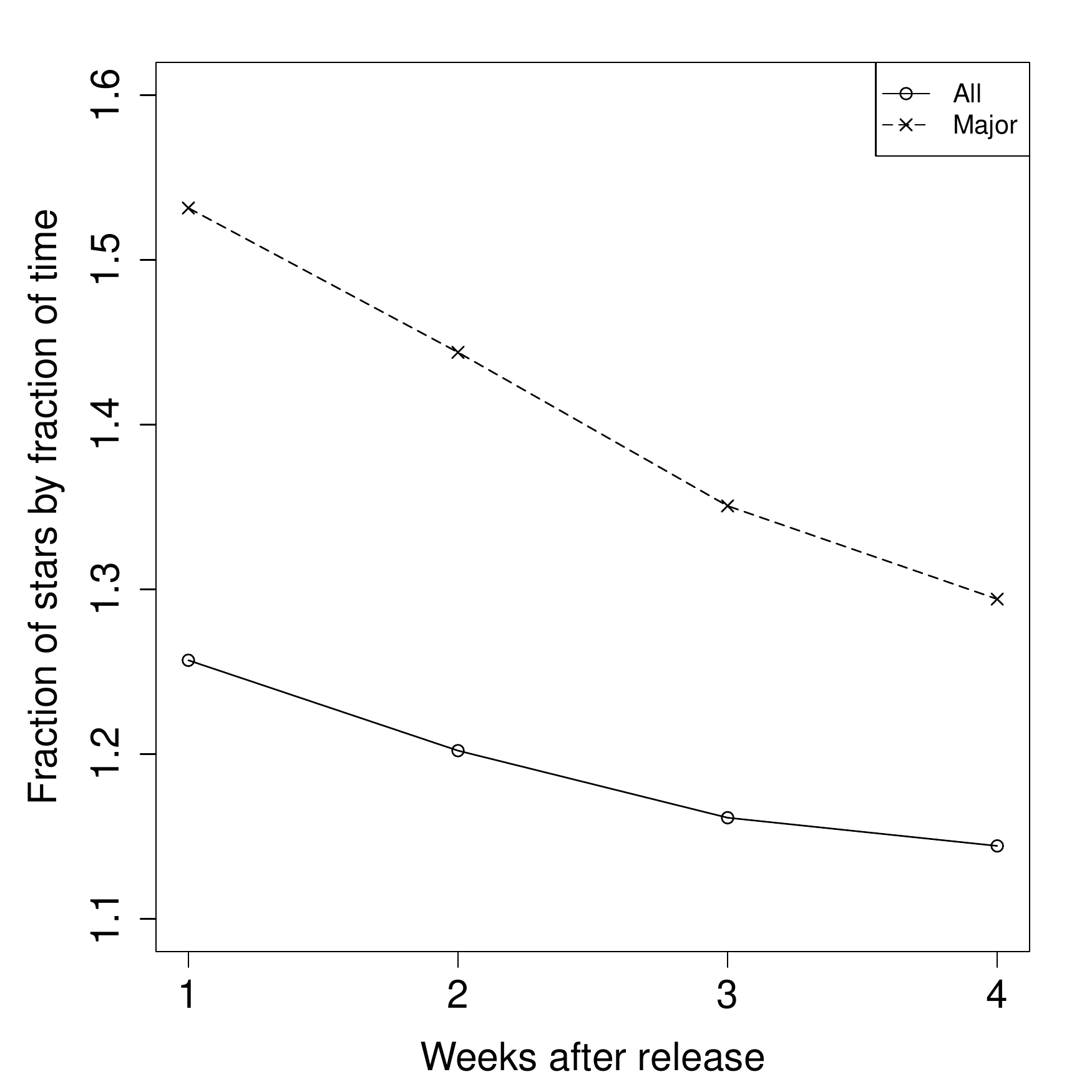}
\caption{Fraction of stars by fraction of time (median values), computed using different time intervals (in weeks)}
\label{fig:releases-values}
\end{figure}

\noindent{\em Summary: There is an acceleration in the number of stars gained just after releases. For example, half of the repositories gain at least 53\% more stars in the week following major releases, than in the other weeks (see Figure~\ref{fig:releases2}). However, because repositories usually have much more weeks without releases than with releases, this phenomenon is not sufficient to generate a major concentration of popularity gain after releases. For example, 75\% of the systems gain at most 4.3\% of their stars in the week following major releases (see Figure~\ref{fig:releases1}).}

\begin{figure*}[!ht]
\centering
\begin{subfigure}[b]{0.25\linewidth}
\includegraphics[width=\linewidth, trim={0 2em 1em 5em}, clip, page=4]{images2/clusters/timeseries.pdf}
\caption{Cluster C1}
\label{fig:timeseries:c1}
\end{subfigure}%
\begin{subfigure}[b]{0.25\linewidth}
\includegraphics[width=\linewidth, trim={0 2em 1em 5em}, clip, page=3]{images2/clusters/timeseries.pdf}
\caption{Cluster C2}
\label{fig:timeseries:c2}
\end{subfigure}%
\begin{subfigure}[b]{0.25\linewidth}
\includegraphics[width=\linewidth, trim={0 2em 1em 5em}, clip, page=2]{images2/clusters/timeseries.pdf}
\caption{Cluster C3}
\label{fig:timeseries:c3}
\end{subfigure}%
\begin{subfigure}[b]{0.25\linewidth}
\includegraphics[width=\linewidth, trim={0 2em 1em 5em}, clip, page=1]{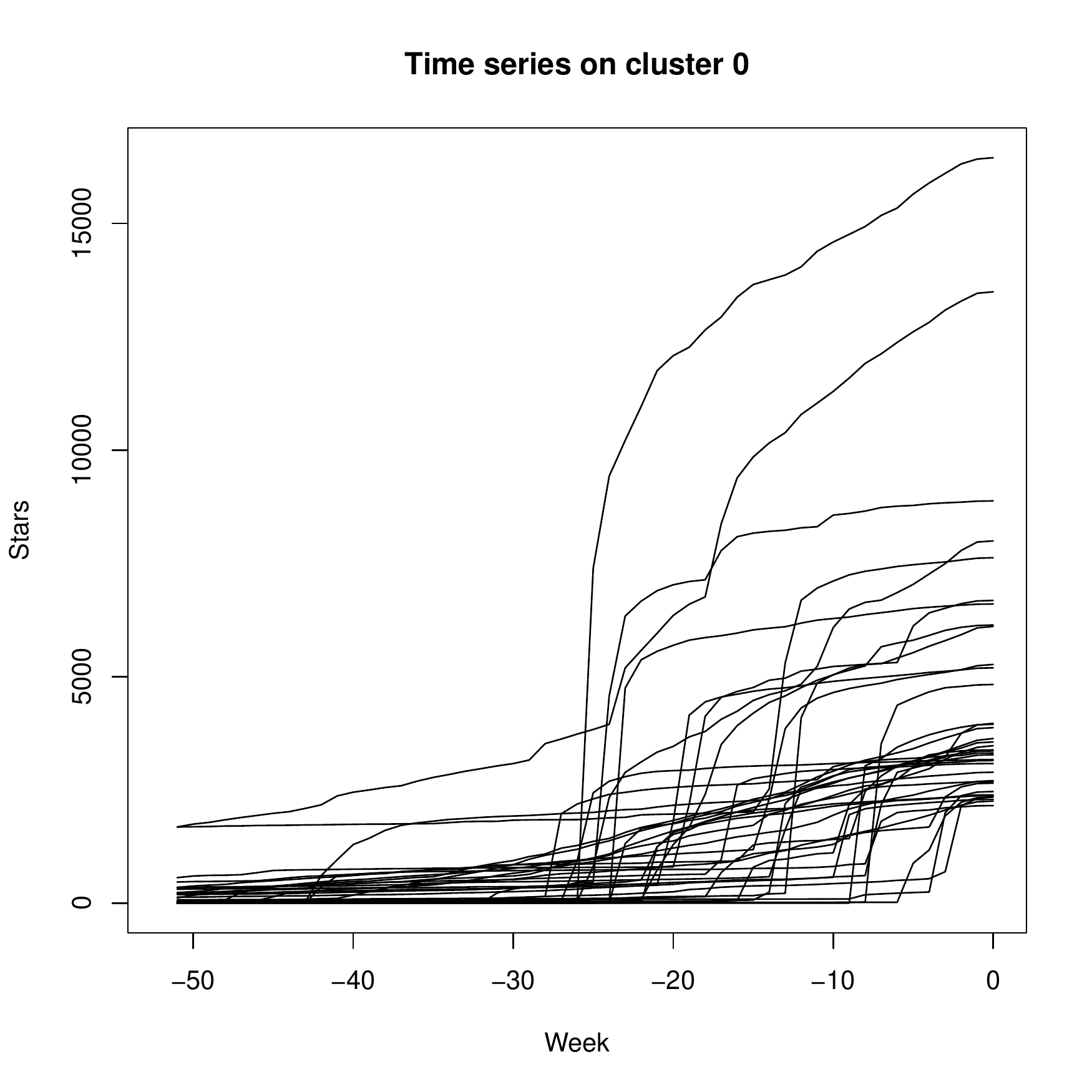}
\caption{Cluster C4}
\label{fig:timeseries:c4}
\end{subfigure}%
\caption{Clusters of time series representing the growth of the number of starts of 2,279 GitHub repositories}
\label{fig:timeseries}
\end{figure*}

\section{Popularity Growth Patterns}
\label{sec-patterns}

In this section, we investigate common patterns of popularity growth concerning the GitHub repositories in our dataset. To this purpose, we use the KSC algorithm~\cite{Yang2011}. This algorithm clusters time series with similar shapes using a metric that is invariant to scaling and shifting.
The algorithm is used in other studies to cluster time series representing the popularity of YouTube videos~\cite{Figueiredo2013} and Twitter~\cite{Lehmann2012}. Like K-means~\cite{hartigan1975}, KSC requires as input the number of clusters $k$.

Because the time series provided as input to KSC must have the same length, we only consider data regarding the last 52 weeks (one year). Due to this restriction, we exclude 216 repositories (8.6\%) that have less than 52 weeks. 
We use the $\beta_{CV}$ heuristic~\cite{Menasce2001} to define the best number $k$ of clusters. $\beta_{CV}$ is defined as the ratio of the coefficient of variation of the intracluster distances and the coefficient of variation of the intercluster distances. The smallest value of $k$ after which the $\beta_{CV}$ ratio remains roughly stable should be selected. This means that new added clusters affect only marginally the intra and intercluster variations~\cite{Figueiredo2014}. In our dataset, the values of $\beta_{CV}$ stabilize for $k=4$ (see Figure~\ref{fig:Bcv}).

\begin{figure}[!h]
\centering
\includegraphics[width=0.65\columnwidth, trim={0 4em 0 5em}, clip]{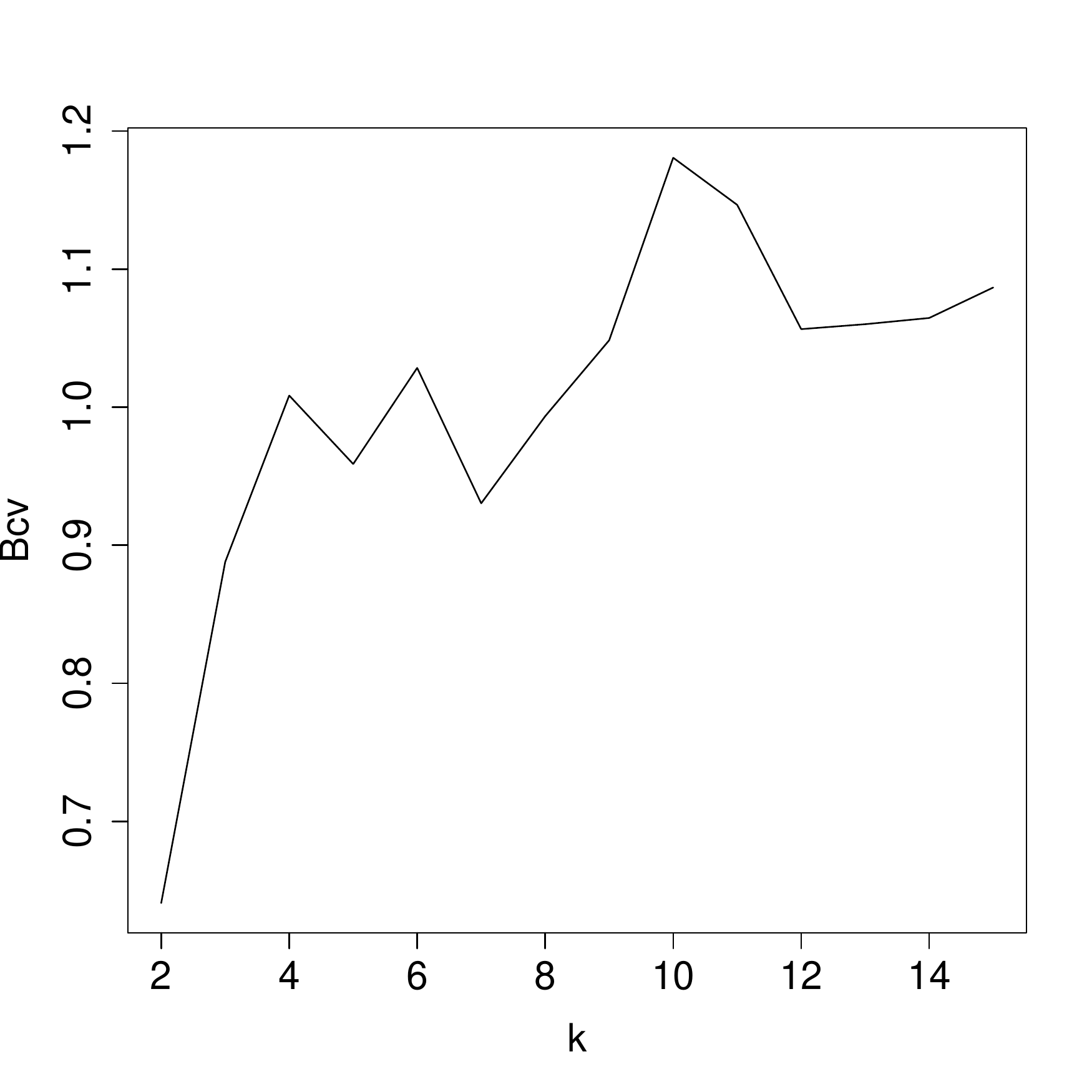}
\caption{$\beta_{CV}$ for $2 \leq k \leq 15$}
\label{fig:Bcv}
\end{figure}

\begin{table*}[!ht]
\centering
\caption{Popularity Growth Patterns}
\label{tab:linear}
\begin{tabular}{@{}lcrrl@{}}
  \toprule
  \multicolumn{1}{c}{\bf Pattern}&\multicolumn{1}{c}{\bf Cluster} & \multicolumn{1}{c}{\bf \# Repositories} & \multicolumn{1}{c}{\bf \% Growth} &  \multicolumn{1}{c}{\bf Top-3 Repositories}\\
  \midrule
  Slow  & C1	& 1,497 (65.7\%) & 27.3  & {\sc jquery/jquery}, {\sc h5bp/html5-boilerplate}, and {\sc meteor/meteor}\\
  Moderate & C2	& 614 (26.9\%) & 94.0 & {\sc facebook/react}, {\sc robbyrussell/oh-my-zsh}, and {\sc airbnb/javascript}\\
	Fast & C3	& 131 (5.7\%) & 469.2 & {\sc atom/electron}, {\sc google/material-design-lite}, and {\sc vuejs/vue} \\
  Viral & C4	& 37 (1.6\%) & 2,673.8 & {\sc nylas/N1}, {\sc letsencrypt/letsencrypt}, and {\sc jwagner/smartcrop.js}\\
  \bottomrule
\end{tabular}
\end{table*}

\subsection{Proposed Growth Patterns}
\label{sec:patterns:trends}

Figure~\ref{fig:timeseries} shows plots with the time series in each cluster. The time series representing the clusters' centroids are presented in Figure~\ref{fig:popularity-trends}.
The time series in clusters C1, C2, and C3 suggest a linear growth, but at different speeds. On the other hand, the series in cluster C4 suggest repositories with a sudden growth on the number of stars. We refer to these clusters as including systems with \emph{Slow}, \emph{Moderate}, \emph{Fast}, and \emph{Viral} Growth, respectively.

\begin{figure}[!ht]
\centering
\includegraphics[width=\columnwidth, trim={0 0 0 0}, clip]{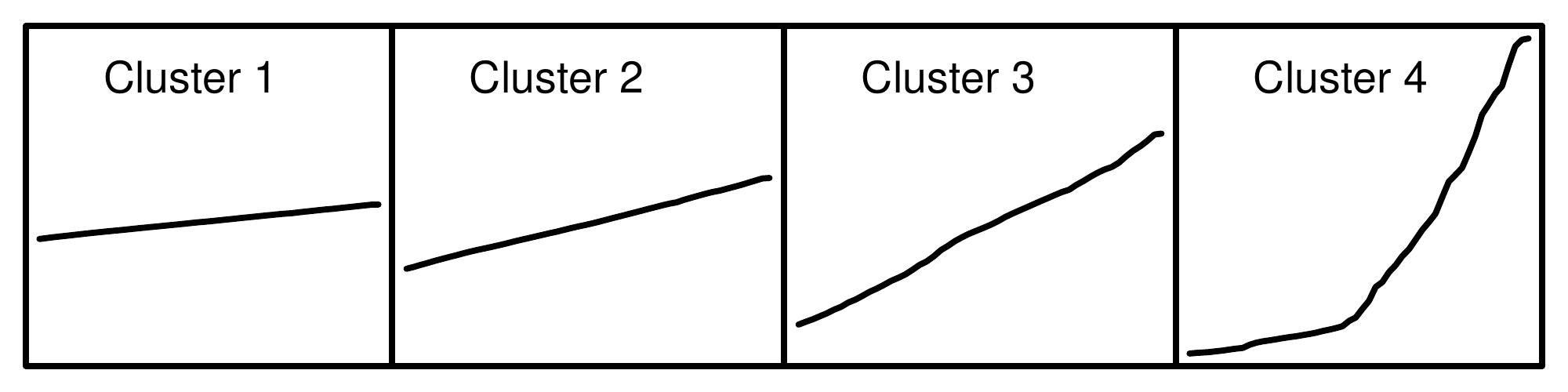}
\caption{Time series representing the centroids of each cluster}
\label{fig:popularity-trends}
\end{figure}

Slow growth is the dominant pattern, including 65.7\% of the repositories in our sample, as presented in Table~\ref{tab:linear}. The table also shows the number of repositories in each cluster and the percentage of stars gained by the cluster's centroids in the period under analysis (52 weeks). The speed in which the stars are gained by repositories on cluster C1 is the lowest one (27.3\% of new stars in one year). Moderate growth is the second pattern with more repositories (26.9\% of the repositories and 94\% of new stars in one year). 5.7\% of the repositories have a fast growth (469.2\% of new stars in the analyzed year).

The last cluster (Viral Growth) describes repositories with a massive growth in their number of stars in a short period of time. It is a less common pattern, including 1.6\% of the repositories. Figure~\ref{fig:viral-examples} shows two examples of systems with a viral growth: {\sc nylas/N1} (an email client, with a peak of more than 7,300 stars in a single week) and {\sc Soundnode/soundnode-app} (a desktop client for SoundCloud, which received almost 1,400 stars in a single week).

\begin{figure}[!ht]
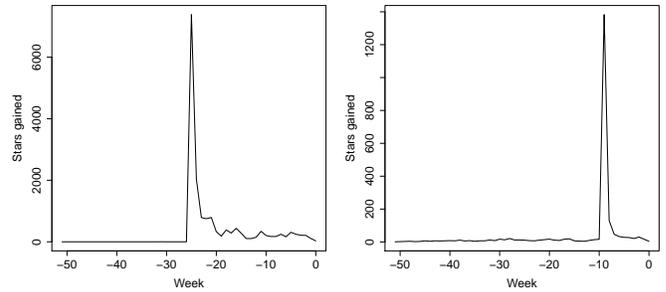

\centering
\begin{subfigure}[t]{0.5\columnwidth}
\centering
\includegraphics[width=1\linewidth, page=18, trim={0 0 0 5em}, clip]{images2/clusters/timeseries_c0.pdf}
\caption{{\sc N1}}
\label{fig:viral-examples:a}
\end{subfigure}%
\begin{subfigure}[t]{0.5\columnwidth}
\centering
\includegraphics[width=1\linewidth, page=11, trim={0 0 0 5em}, clip]{images2/clusters/timeseries_c0.pdf}
\caption{{\sc Soundnode}}
\label{fig:viral-examples:b}
\end{subfigure}%
\caption{Examples of viral growth}
\label{fig:viral-examples}
\end{figure}

\begin{figure*}[!ht]
\centering
\includegraphics[width=\textwidth, trim={0 2em 0 0em}, clip]{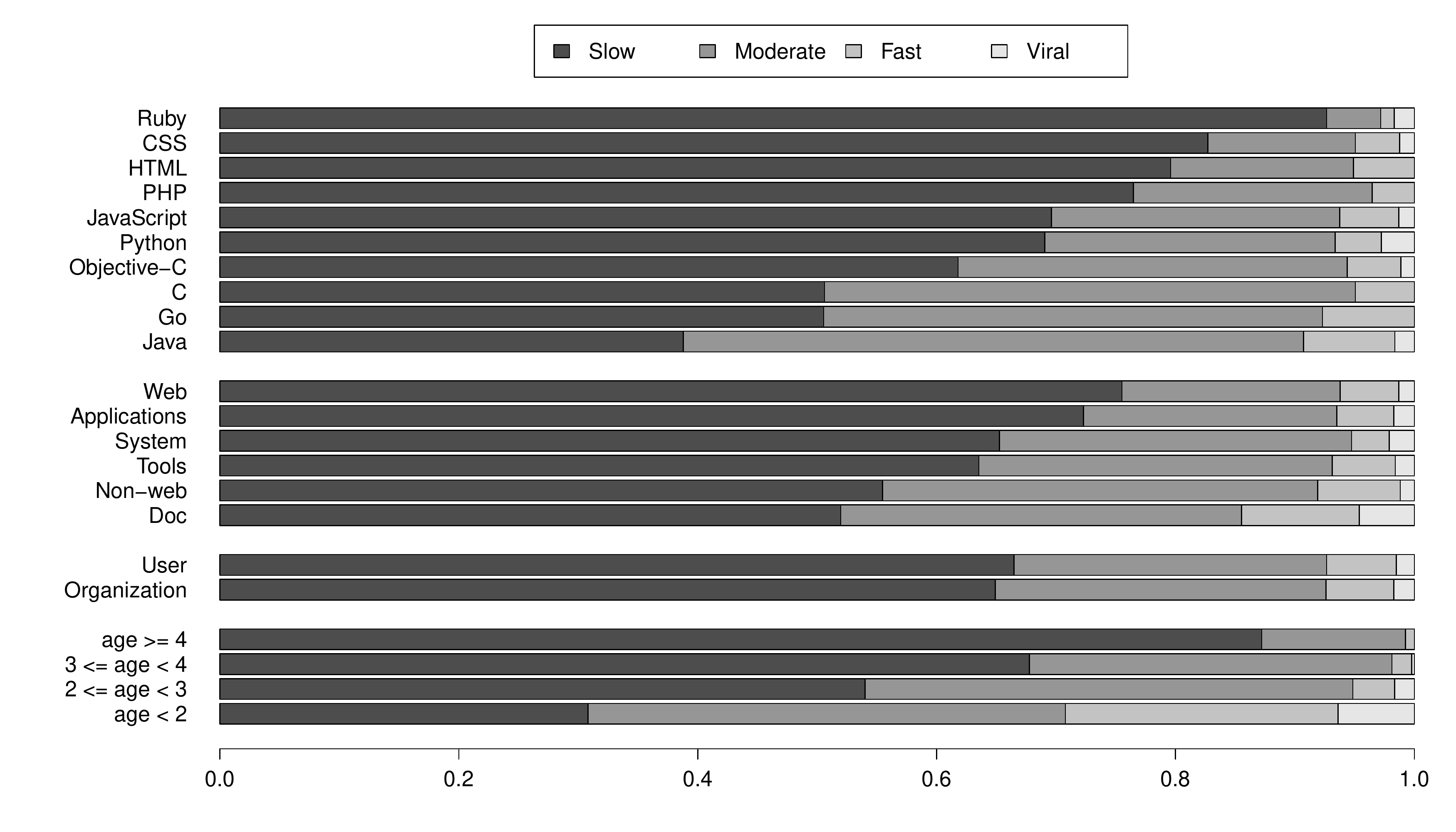}
\caption{Percentage of systems following the proposed growth patterns, for the most popular programming languages, application domains, and repository owners}
\label{fig:stackedbars}
\end{figure*}

\subsection{Growth Patterns vs Repositories Properties}

Figure~\ref{fig:stackedbars} shows the percentage of systems following the proposed growth patterns, for the top-10
programming languages in number of repositories, application domains, repository owners, and age.
The three languages with the highest percentage of systems with slow growth
are Ruby (92\%), CSS (82\%), and HTML (79\%).
By contrast, the languages with the highest percentage of systems
with fast growth are Go (7.6\%) and Java (7.6\%). Go is a new
language that is attracting a lot of interest.\footnote{https://www.thoughtworks.com/radar/languages-and-frameworks/go-language, verified on 04/07/2016.}
Regarding Java, 61 out of 95 repositories with fast growth are Android applications.
JavaScript is the language with the highest number of repositories with
viral growth (10 repositories), followed by C++ (5 repositories) and Python (5 repositories). In relative
terms, 2.7\% of the Python systems have a viral growth, followed
by 1.6\% of the systems implemented in Ruby.
When we group the systems by application domain, 75\% of the web libraries and
frameworks have a slow growth.
Interestingly, the two domains with the highest
percentage of systems following a fast growth are documentation (9.8\%)
and non-web libraries and frameworks (6.9\%).
Regarding the repository owners, there is no substantial difference between
users and organizations. For slow growth, the percentage
of systems is 66.4\% and 64.9\%, for users and organizations, respectively.
For fast growth, the percentage is 5.8\% and 5.6\%, respectively.
Finally, the last bars in Figure~\ref{fig:stackedbars} show that old repositories tend to
present a slow growth. The percentage of such repositories ranges from 30.8\%
(age $<$ 2 years) to  87.2\% (age $\geq$ 4 years).

As mentioned,  we found a high percentage of web frameworks and libraries---especially the ones implemented in Ruby, CSS, and HTML---with a slow growth. We hypothesize two main reasons to explain this result. First,
web libraries and frameworks are the dominant applications in our dataset of popular applications (837 repositories, 33\%). This implies in a high competition, with many systems disputing the same users. For example,
we found a list of JavaScript MVC-based frameworks with slow growth, including systems like
{\sc knockout/knockout}, {\sc spine/spine}, {\sc quirkey/sammy}, and {\sc sproutcore/sproutcore}.
These systems have to compete with ``blockbusters'', like {\sc angular/angular.js},
which is certainly a challenging task.  The second reason is that there are many highly popular web frameworks and libraries in our dataset. For example, among the top-10\% repositories in number of stars, 42.8\% are web libraries and frameworks. We cannot assume that these systems will present the same growth rates of less popular ones. For instance, if {\sc angular/angular.js} starts to grow at 469.2\% per year (the growth rate observed for the centroid of the repositories with fast growth) it will have almost 1.5M stars in two years.

\section{Feedback from Developers}
\label{sec:survey}

We contacted the main developers of some GitHub repositories to clarify the results presented in the previous sections. Specifically, we surveyed developers about three themes: (a) the impact on popularity of repositories owned by users (Section~\ref{sec:survey-user}); (b) the main characteristics of successful releases (Section~\ref{sec:survey-releases}); (c) the reasons for the peaks of popularity observed in systems with viral growth (Section~\ref{sec:survey-viral}). The surveys were performed by means of follow-up emails.

\subsection{Impact on Popularity of Repositories Owned by Users}
\label{sec:survey-user}

In Section~\ref{sec:results} (RQ \#1), we found that repositories owned by organizations are more popular than the ones owned by individuals. For example, among the top-100 most popular repositories, only 30 repositories are owned by users. The developers of  17 of such systems have a public mail address in their GitHub profile. We sent a short survey to these developers and received responses from five of them (29.8\%). In this survey, we asked two questions. First, we asked the developers about possible plans to migrate their repositories to an organization account. All developers answered negatively this question. Two developers mentioned they want to explicitly appear as the repository owner, like in this answer: \\[-0.3cm]

\noindent {\em ``I worked hard to create the project, and having it under my personal username is necessary to have proper credit for it.''} \\[-0.3cm]

To complement the first question, we asked the developers if they agree that migrating the repositories to an organization account would help to attract more users. Four developers (80\%) answered negatively to this question and only one participant provided the following answer:\\[-0.3cm]

\noindent {\em``It depends on what organization it is. If it's a well known org I'm sure it helps, otherwise I don't think it makes a difference.''} \\[-0.3cm]

Therefore, although it seems ``easier'' to organizations to reach the top positions of GitHub popularity ranking, some systems owned by individual developers also reach these positions. These developers usually do not want to move to organizational accounts, basically to keep full control and credit for their repositories.

\subsection{Characteristics of Successful Releases}
\label{sec:survey-releases}

To reveal the characteristics of the most successful releases in our dataset (see RQ \#4, Section~\ref{sec:results}), we perform a survey with the main developers of 60 releases with the highest fraction of stars gained on the week after the release (and whose developers have a public mail address on their GitHub profile). We received answers from 25 developers, which corresponds to a response ratio of 41.6\%. First, we asked the developers about the type of features implemented in these releases. As presented in Table~\ref{tab:question1}, the releases usually include both functional and non-functional requirements (14 answers), followed by releases with mostly functional requirements (9 answers). We did not receive answers about releases including non-functional requirements. Two developers provide other types of answers (``complete rewrite'' and ``maintenance release'', respectively).

\begin{table}[!ht]
\centering
\caption{Features implemented in successful releases}
\label{tab:question1}
\begin{tabular}{@{}lrl@{}}
\toprule
Features & \multicolumn{2}{c}{Answers}  \\
\midrule
Both functional and non-functional & 14 & \sbar{14}{25}  \\
Mostly functional & 9 & \sbar{9}{25}  \\
Other answers & 2 & \sbar{2}{25}  \\
Mostly non-functional & 0 & \sbar{0}{25}  \\
\bottomrule
\end{tabular}
\end{table}

We also asked the developers to explain how the features implemented in these releases were selected (answers including multiple items are possible in this question). As presented in Table~\ref{tab:question2}, the features usually come from ideas of the repository' maintainers (23 answers) and from user's suggestions (11 answers).

\begin{table}[!ht]
\centering
\caption{How the features are selected?}
\label{tab:question2}
\begin{tabular}{@{}lrl@{}}
\toprule
Features selected from & \multicolumn{2}{c}{Answers}  \\
\midrule
Ideas of the repository maintainers	& 23 & \sbar{23}{25}  \\
Users suggestions	& 11 & \sbar{11}{25}  \\
Features of similar projects	& 6	& \sbar{6}{25}  \\
Other answers	& 3	& \sbar{3}{25}  \\
\bottomrule
\end{tabular}
\end{table}

\subsection{Reasons for Viral Growth}
\label{sec:survey-viral}

To expose the reasons for viral growth, we sent a message to the main developer of 22 systems with viral growth and who have a public mail address on their GitHub profile. In the message, we asked the developers to  explain the peaks observed in the number of stars of their repositories. We received answers from 14 developers, which corresponds to a response ratio of 63\%. As presented in Table~\ref{tab:survey-viral}, 11 developers (78.5\%) linked the peaks to posts in social media sites, mostly Hacker News.\footnote{https://news.ycombinator.com/} For example, we received the following answer:\\[-0.3cm]

\noindent{\em \aspas{I posted about this project on HackerNews. It quickly got a lot of attention and remained on the front page of HackerNews (a very high traffic tech site) for over 24 hours. It subsequently made it onto the github.com/explore as one of the top starred repositories for around a week. Because the repo was highlighted in these two high-profile locations for so much time, it received an incredible amount of traffic, which translated to a considerable number of stars.}}\\[-0.3cm]



\begin{table}[!ht]
\centering
\caption{Sources of popularity}
\label{tab:survey-viral}
\begin{tabular}{@{}lrl@{}}
\toprule
Source & \multicolumn{2}{c}{Answers}  \\
\midrule
Social media sites (e.g., HackerNews) & 11 & \sbar{11}{14}  \\
Blogs and news sites (e.g., infoq.com)       & 3  & \sbar{3}{14}  \\
Other answers (e.g., private mailing list)      & 4  & \sbar{4}{14}  \\
\bottomrule
\end{tabular}
\end{table}

\section{Threats to Validity}
\label{sec:threats}

\noindent{\em Number of stars as a proxy for popularity:} In the paper, we consider that stars are proxies for a project popularity, as common in studies about the popularity of social media content~\cite{Ahmed2013, Figueiredo2014,Lehmann2012, ma2013predicting}. However, a developer can star a repository for other reasons, for example, when she in fact finds problems in the system and wants to create a bookmark for later access and analysis.\\[-0.25cm]

\noindent{\em Dataset.} GitHub has millions of repositories.
We build our dataset by collecting the top-2,500 repositories with more stars, which represents a small fraction in comparison to the GitHub's universe. However, our goal is exactly  to investigate the popularity of the most starred repositories.
Furthermore, most GitHub repositories are forks and have very low activity~\cite{Kalliamvakou2014, Kalliamvakou2015}.\\[-0.25cm]

\noindent{\em Application domains.} Because GitHub does not classify the hosted applications in domains, we performed this classification manually. Therefore, it is subjected to errors and inaccuracies. To mitigate this threat, the dubious classification decisions were discussed by two paper's authors.\\[-0.25cm]

\noindent{\em Growth patterns}. The selection of the number of clusters is a key parameter in algorithms like KSC. To mitigate this threat, we employed a heuristic that considers the intra/intercluster distance variations~\cite{Menasce2001}. Furthermore, the analysis of growth patterns was based on the stars obtained on the last year. The stars before this period are not considered, since the KSC algorithm requires time series with the same length. \\[-0.25cm]


\section{Related Work}
\label{sec:related-work}

Several studies examine the relationship between popularity of mobile apps and their code properties~\cite{Datta2013, Fu2013, LinaresVasquez2013, Ruiz2014, lee2014, Tian2015, Guerrouj2015, Palomba2015, Corral2015}.
Yuan et al.~investigate 28 factors along eight dimensions to understand how high-rated Android applications are different from low-rated ones~\cite{Tian2015}. Their result shows that external factors, like number of promotional images, are the most influential factors.
Guerrouj and Baysal explore the relationships between mobile apps' success and API quality~\cite{Guerrouj2016}. They found that changes and bugs in API methods are not strong predictors of apps' popularity.
Ruiz et al.~examine the relationship between the number of ad libraries and app's user ratings~\cite{Ruiz2014}. Their results show that there is no relationship between the number of ad libraries in an app and its rating.
Linares-V{\'a}squez et al.~investigate how the fault- and change-proneness of Android API elements relate to applications' lack of success~\cite{LinaresVasquez2013}. They state that making heavy use of fault- and change-prone APIs can negatively impact the success of these apps.

Other studies examine source code repositories in order to understand what makes a project popular.
Weber and Luo attempt to differentiate popular and unpopular Python projects on GitHub using machine learning techniques~\cite{Weber2014}. They found that in-code features are more important than author metadata features.
Zho et al.~study the frequency of folders used by 140 thousands GitHub projects and the results suggest that the use of standard folders (e.g., doc, test, examples) may have an impact on project popularity~\cite{Zhu2014}.
Bissyande et al.~analyze the popularity, interoperability, and impact of various programming languages, using a dataset of 100K open source software projects~\cite{Bissyande2013}.
Aggarwal et al.~study the effect of social interactions on GitHub projects' documentation~\cite{Aggarwal2014}. They conclude that  popular projects tend to attract more documentation collaborators.
By analyzing usage of Java APIs, Mileva states that popularity trend is a method for displaying the users preferences and for predicting their future~\cite{Mileva2012}.

Finally, other studies analyze the relationship between popularity and software quality.
Sajnani et. al.~ study the relationship between component popularity and component quality in Maven~\cite{Sajnani2014}, finding that, in most cases, there is no correlation.
Capra et. al.~evaluate the effect of firms' participation on communities of open source projects and conclude that firms' involvement improves the popularity, but leads to lower software quality~\cite{Capra2011144}.

To our knowledge, we are the first study to track popularity over time on social code sharing sites, like GitHub.
However, there are similar studies in other contexts, like App Stores~\cite{Guerrouj2015}, video sharing sites~\cite{Chatzopoulou2010}, and social platforms~\cite{ma2013predicting}.
Chatzopoulou et al.~\cite{Chatzopoulou2010} analyze popularity of Youtube videos by looking at properties and patterns metrics. They report that many of the popularity metrics are highly correlated. In our study, we also report correlations between stars and other popularity metrics (e.g., forks).
Lehmann et al.~\cite{Lehmann2012} analyze popularity peaks of hashtags in Twitter. They found four usage patterns restricted to a two-week period centered on the peak time whereas the popularity patterns presented in this study are based on the last year data.

\section{Conclusion}
\label{sec-conclusion}

In this paper, we first studied the popularity of GitHub repositories aiming to answer four research questions.
We concluded  that three most common domains on GitHub are web libraries and frameworks, non-web libraries and frameworks, and software tools.
However, the three domains whose repositories have more stars are systems software, web libraries and frameworks, and documentation.
Additionally, we found that repositories owned by organizations are more popular than the ones owned by individuals (RQ \#1).
We also reported the existence of a strong correlation between stars and forks, a weak correlation between stars and commits, and  a weak correlation between stars and contributors (RQ \#2), confirming the importance of a large base of contributors to the success of open source software~\cite{Mockus2002}.
We concluded that repositories have a tendency to receive more stars right after their first public release.
After this period, for half of the repositories the growth rate tends to stabilize (RQ \#3).
In other words, bursts of popularity do not explain the popularity growth of most repositories.
We showed that there is an acceleration in the number of stars gained just after releases (RQ \#4), which
confirms the importance of developers constantly evolving and improving their systems.

We identified four patterns of popularity growth, which were derived after
clustering the time series that describe the number of stars of the systems in our dataset.
We found that slow growth is the most common pattern (65.7\%) and that very few systems present a viral behavior (1.6\%).
Slow growth is more common in case of overpopulated application domains (as web libraries and frameworks) and for old repositories.


As future work, we plan to investigate repositories that are not popular yet and compare them with the popular ones.
We also plan to correlate repository and language popularity to provide relative measures of popularity.
For example, if we restrict the analysis to developers from a given language, a Scala repository can be considered more popular than a JavaScript one, although having less stars.
Moreover, we plan to investigate models for predicting software popularity, which can be used for example to warn developers when signs of stagnation are detected in their repositories.

\section*{Acknowledgments}

\noindent This research is supported by FAPEMIG and CNPq.

\bibliographystyle{IEEEtran}
\bibliography{IEEEabrv,references}

\end{document}